\documentclass{article}

\usepackage{arxiv}

\usepackage[utf8]{inputenc} 
\usepackage[T1]{fontenc}    
\usepackage{hyperref}       
\usepackage{url}            
\usepackage{booktabs}       
\usepackage{amsfonts}       
\usepackage{nicefrac}       
\usepackage{microtype}      
\usepackage{lipsum}		
\usepackage{graphicx}
\usepackage{natbib}
\usepackage{doi}

\usepackage[version=3]{mhchem} 
\usepackage{xcolor}
\usepackage{xr-hyper}
\usepackage{hyperref}
\usepackage{amssymb}
\usepackage{caption}
\usepackage{subcaption}

\title{Gibbs-Helmholtz Graph Neural Network: \\ capturing the temperature dependency of  activity coefficients at infinite dilution}

\date{} 					

\author{ 
    Edgar Ivan Sanchez Medina \\
	Chair for Process Systems Engineering\\
	Otto-von-Guericke University\\
	Universitätsplatz 2, 39106, Magdeburg, Germany \\
	\texttt{sanchez@mpi-magdeburg.mpg.de} \\
	\And
	Steffen Linke \\
	Chair for Process Systems Engineering\\
	Otto-von-Guericke University\\
	Universitätsplatz 2, 39106, Magdeburg, Germany \\
    \texttt{linke@mpi-magdeburg.mpg.de} \\
	\And 
    Martin Stoll \\
    Chair of Scientific Computing\\
	Technische Universität Chemnitz\\
	09107, Chemnitz, Germany \\
    \texttt{martin.stoll@mathematik.tu-chemnitz.de} \\
	\And 
    Kai Sundmacher \\
    Chair for Process Systems Engineering\\
	Otto-von-Guericke University\\
	Universitätsplatz 2, 39106, Magdeburg, Germany \\
    \hfill \break \\
    Process Systems Engineering\\
	Max Planck Institute for \\ Dynamics of Complex Technical Systems\\
	Sandtorstr. 1, 39106, Magdeburg, Germany \\
    \texttt{sundmacher@mpi-magdeburg.mpg.de} \\
}

\renewcommand{\shorttitle}{\textit{arXiv} Template}

\hypersetup{
pdftitle={A template for the arxiv style},
pdfsubject={q-bio.NC, q-bio.QM},
pdfauthor={David S.~Hippocampus, Elias D.~Striatum},
pdfkeywords={First keyword, Second keyword, More},
}

\begin{document}
\maketitle

\begin{abstract}
	The accurate prediction of physicochemical properties of chemical compounds in mixtures (such as the activity coefficient at infinite dilution $\gamma_{ij}^\infty$) is essential for developing novel and more sustainable chemical processes. In this work, we analyze the performance of previously-proposed GNN-based models for the prediction of $\gamma_{ij}^\infty$, and compare them with several mechanistic models in a series of 9 isothermal studies. Moreover, we develop the Gibbs-Helmholtz Graph Neural Network (GH-GNN) model for predicting $\ln \gamma_{ij}^\infty$ of molecular systems at different temperatures. Our method combines the simplicity of a Gibbs-Helmholtz-derived expression with a series of graph neural networks that incorporate explicit molecular and intermolecular descriptors for capturing dispersion and hydrogen bonding effects. We have trained this model using experimentally determined $\ln \gamma_{ij}^\infty$ data of 40,219 binary-systems involving 1032 solutes and 866 solvents, overall showing superior performance compared to the popular UNIFAC-Dortmund model. We analyze the performance of GH-GNN for continuous and discrete inter/extrapolation and give indications for the model's applicability domain and expected accuracy. In general, GH-GNN is able to produce accurate predictions for extrapolated binary-systems if at least 25 systems with the same combination of solute-solvent chemical classes are contained in the training set and a similarity indicator above 0.35 is also present. This model and its applicability domain recommendations have been made open-source at \url{https://github.com/edgarsmdn/GH-GNN}. 
\end{abstract}


\section{Introduction}
The current efforts of switching the foundation of the chemical industry from fossil-based resources to more sustainable options require the design of novel chemical processes that involve new molecules and materials. Perhaps, the most important type of processes to be newly designed and optimized towards this goal are separation processes. These type of processes already account for 10-15\% \citep{sholl2016seven} of the world's energy consumption, and  constitute 40-50\% \citep{kiss2016separation} of the total costs in the largest chemical plants worldwide. 

If novel and more sustainable separation processes are to be designed, the availability for solid thermodynamic data of promising chemicals is of great importance \citep{mcbride2020hybrid}. However, when considering the enormous chemical space of all synthesizable molecules, our limited experimental capacity for measuring thermodynamic data for all possible compounds becomes evident. This already massive chemical space is just a subset of an orders of magnitude larger state space when considering all possible mixtures at different composition, pressure and temperature conditions. For instance, for capturing the entire phase equilibrium of a single 10-component system at constant pressure the collection of the necessary data points to be measured is estimated to last around 37 years \citep{gmehling2019chemical} if one assumes constant pressure and 10 \%mol steps. These clear experimental limitations have motivated the development and use  of thermodynamic predictive methods over decades \citep{gmehling2019chemical}.

The modeling and design of separation processes relies on the accurate prediction of phase equilibria. In non-ideal liquid mixtures, phase equilibrium is governed by the activity coefficient $\gamma_i$ which measures the degree of deviation from an ideal solution that the component $i$ has due to inter- and intramolecular interactions. In chemical engineering, binary activity coefficients at infinite dilution $\gamma_{ij}^\infty$ (where $i$ and $j$ denote the solute and solvent, respectively) are of special interest for several reasons. First, in high purity regimes, the real behavior of the mixture cannot be reliably extrapolated from finite dilution data. This is usually the scenario when impurities, catalysts or by-products are present in very low concentrations (e.g., pollutants in waste water streams \citep{sandler1996infinite}).  Second, usually the selection of a suitable entrainer in extractive separation processes involves the analysis of $\gamma_{ij}^\infty$ values because they provide a good initial estimation of the performance of the achievable separation and elucidate potential separation problems such as azeotropic points and limited miscibility \citep{krummen2000measurement, harten2020software}. Third, it is possible to calculate activity coefficients over the whole composition range by using the binary $\gamma_{ij}^\infty$ values of all the species involved in the mixture (e.g., by using two-parameter activity coefficient models \citep{brouwer2021solvent}). 

Several models have been developed that are suitable to predict $\gamma_{ij}^\infty$ values, and they can be broadly classified into phenomenological or mechanistic models (both referred to as mechanistic models), and the models that are based on machine learning techniques. The most commonly used mechanistic models are UNIFAC-Dortmund \citep{gmehling1998modified, lohmann2001unifac} which is based on functional molecular groups and binary group-interaction parameters fitted from experimental data, and COSMO-RS \citep{eckert2002fast} which relies on statistical thermodynamics and quantum chemistry calculations. Probably, the main reason why these two models are so popular is not their particular accuracy, but that they can be applied to a very broad range of molecules. While the applicability of UNIFAC-Dortmund is only limited by the feasibility of the molecule's fragmentation into the pre-defined UNIFAC groups and the availability of all necessary binary interaction parameters, COSMO-RS is, in principle, able to predict $\gamma_{ij}^\infty$ for any solute-solvent combination as soon as the necessary molecular surface charge distribution from DFT calculations are available. However, other models such as MOSCED \citep{lazzaroni2005revision} are able to provide more accurate $\gamma_{ij}^\infty$ predictions than the two previously mentioned models for several types of systems \citep{brouwer2019model}. The problem is that the space of molecules that models like MOSCED can predict is much more limited compared to the popular UNIFAC-Dortmund and COSMO-RS methods. The main difference of MOSCED, compared to UNIFAC-Dortmund, is that it incorporates explicit molecular descriptors that capture different solute-solvent interactions (i.e., dispersion, induction, polarity and hydrogen bonding interactions). Despite of the efforts that have been made on developing mechanistic models that predict $\gamma_{ij}^\infty$ as accurately as possible, many unexplained deviations from experimental values still exist \citep{brouwer2019model}.

With the recent advances in machine learning methods along with the increase of computational power, many data-driven approaches have been recently investigated for predicting $\gamma_{ij}^\infty$. Early attempts used the classical QSPR approach based on pre-calculated molecular descriptors and regression techniques \citep{behrooz2017prediction, ajmani2008characterization, paduszynski2016silico}. These methods were trained on limited data which makes their generalization ability questionable.  In the last years, matrix completion methods have been investigated for the prediction of $\gamma_{ij}^\infty$ both at constant \citep{jirasek2020hybridizing, jirasek2020machine} and varying \citep{damay2021predicting, tanprediction, chen2021neural} temperature. However, the predictions of this type of models are naturally limited by the size of the solute-solvent matrix on which they were trained. Despite of this being a clear limitation of the method for exploring the entire chemical space, it is also advantageous from the perspective of having a clear applicability domain for accurate predictions. Another method based on transformers and natural language processing called  SMILES-to-Properties-Transformer (SPT) have been recently developed by  \cite{winter2022smile} achieving notable results. Their approach involves a pre-training step on a very large dataset obtained from COSMO-RS predictions. By employing neural networks and attention mechanisms the model learns to map the SMILES of the solute and solvent to the $\gamma_{ij}^\infty$ value. The temperature dependency is included as part of the learned sequence embedding. In our previous work \citep{D1DD00037C}, we introduce a method based on graph neural networks (GNNs) for predicting $\gamma_{ij}^\infty$ at standard temperature achieving lower errors than UNIFAC-Dortmund and COSMO-RS in the dataset studied. Hybrid graph neural network models were also trained on the mechanistic models' residuals overall reducing the errors even further. However, the extrapolation performance of these GNN-based models was not analyzed. Recently, \cite{qin2022capturing} developed SolvGNN, a graph neural network framework that captures hydrogen-bonding interactions explicitly and allows for activity coefficient predictions at various compositions. However, their implementation of SolvGNN was trained on isothermal simulated data obtained from COSMO-RS which inherently bounds its actual prediction accuracy to the accuracy of COSMO-RS itself. More recently, the temperature dependency has been included into graph neural network models for the predictions of $\gamma_{ij}^\infty$  in systems involving ionic-liquids \citep{rittig2022graph}. 

In the present work, we use experimentally measured $\gamma_{ij}^\infty$ data collected in the DECHEMA Chemistry Data Series \citep{dechema}. The cleaned experimental dataset that we used includes data of 866 solvents and 1032 solutes which is considerably larger compared to the experimental data used in the above mentioned related works (i.e., approximately twice the number of solutes and solvents considered in other works). This allow us to test our model in a larger chemical space potentially increasing its applicability domain.  We first analyze the performance of the GNN architecture proposed in our previous work \citep{D1DD00037C} and the SolvGNN architecture introduced by  \cite{qin2022capturing} for predicting $\gamma_{ij}^\infty$ in a set consisting of 9 isothermal subsets. We compare both methods to the mechanistic models UNIFAC-Dortmund, COSMO-RS and MOSCED. Then, we proposed a physics-based GNN architecture (GH-GNN) for capturing the temperature dependency of $\gamma_{ij}^\infty$ based on a relation derived from the Gibbs-Helmholtz equation and employing ideas from the SolvGNN and MOSCED models. Following this, we study the continuous inter/extrapolation capabilities of GH-GNN to different temperatures, and its performance on discrete inter/extrapolation to different solute-solvent combinations showing that GH-GNN is able to accurately predict $\gamma_{ij}^\infty$ over a large range of temperatures and solute-solvent combinations. Moreover, we discuss aspects of the applicability domain of GH-GNN which is an important aspect often bypassed in most machine learning methodologies proposed in the literature. The model has been made open-source at \url{https://github.com/edgarsmdn/GH-GNN}.      

This paper is structured as follows: we first describe the experimental dataset and the data cleaning process that we used. Then, we define the molecular and mixture graphs and the GNN architectures used for our isothermal and temperature-dependent studies. Afterwards, the results for these studies are presented and the performances for continuous and discrete inter/extrapolation are discussed. Finally, we conclude our work and suggest some future research directions.

\section{Methods}
\subsection{Data sources}

The data used in the present study is a subset of the data collected on the DECHEMA Chemistry Data Series Vol. IX \citep{dechema}. This collection is among the largest experimental data sets for $\gamma_{ij}^\infty$ that has been gathered and curated over the years. Despite this database being not open-source, the use of it allows for the open-source release of the models that are developed from it (such as in the present work). This is in contrast to the legal restrictions of releasing complete open-source models constructed from similar-size experimental data collections (e.g., Dortmund Data Bank (DDB) \citep{DDB}). 

The complete DECHEMA data collection consists of $\gamma_{ij}^\infty$ for binary systems measured mainly by the following experimental methods: Gas-liquid chromatography, derived from solubility data, dilutor technique, static method, ebulliometry and other techniques such as liquid-liquid chromatography and the Rayleigh distillation method. An excellent review on such techniques is provided by \cite{dohnal200514}. In this data collection, only $\gamma_{ij}^\infty$ values determined by dedicated experimental techniques were included, as values extrapolated from phase equilibrium measurements at finite dilution tend to be inaccurate \citep{sandler1996infinite}. See section S1 in the Supporting Information for a list of all experimental techniques involved and the proportion of data points measured by each experimental technique.

Despite of having a broad overview of the experimental techniques used to measure the collected $\gamma_{ij}^\infty$, the vast majority of works in the scientific literature reporting experimental measurements of $\gamma_{ij}^\infty$ do not report confidence intervals. However, some general uncertainty estimations can still be found in the literature. For instance,  \cite{damay2021predicting}, have stated that typical absolute experimental $\ln \gamma_{ij}^\infty$ uncertainties range from 0.1 to 0.2. By contrast, some authors have reported a relative experimental uncertainty between 1\% and 6\% on $\gamma_{ij}^\infty$ \citep{lerol1977accurate, domanska2019ammonium, vrbka2016limiting, marcinkowski2020measurements}. As a result, \cite{brouwer2021trends} have estimated a minimum relative uncertainty of  5\% for the data they have collected \citep{brouwer2021trends}, which is in-line with the overall uncertainties calculated elsewhere \citep{bahadur2014measurement, dohnal200514}. 

\subsection{Data cleaning}

Roughly 91\% of the binary-systems included in the DECHEMA Chemistry Data Series Vol. IX \citep{dechema} correspond to systems involving only molecular compounds. The rest of the systems involve ionic liquids. In this work, we have only considered binary molecular systems, but the proposed method can be extended to train models on systems including ionic-liquids. A recent work\citep{rittig2022graph} has precisely covered this type of systems in the context of GNNs for $\gamma_{ij}^\infty$ prediction. Whenever multiple measurements of the same binary systems at the same temperature were found in the DECHEMA Chemistry Data Series Vol. IX \citep{dechema} these were averaged to obtain a single value. Also, compounds with non-identified isomeric configurations or ambiguous SMILES identification (e.g., commercial solvents like Genosorb 300) were excluded from the dataset. The resulting dataset (referred to as the DECHEMA dataset) covers 40,219 data points which include 866 solvents and 1032 solutes. This number of solvents and solutes result in 893,712 possible binary combinations out of which only 1.64\% (14,663 binary-systems) were actually measured. 

This DECHEMA dataset used in the present work is considerably larger (both in terms of the number of chemical species covered and the number of experimental data points gathered) than the data sets used by recent machine learning approaches based on matrix completion methods \citep{damay2021predicting, tanprediction} and natural language processing \citep{winter2022smile}. Moreover, despite covering more experimental data points the density of the observed entries of the solute-solvent matrix is considerably lower than in the recent works above mentioned. For instance, the matrix completion method of \cite{damay2021predicting} covers only 414 solvents and 378 solutes (52\% and 63\% less solvents and solutes than in the present work, respectively) with a total of 7107 observed binary systems resulting in a more populated matrix (4.54\% of observed points of the complete matrix compared to our 1.64\%). The matrix of chemical species covered is even smaller in the case of the method presented by \cite{winter2022smile} covering only 349 solvents and 373 solutes (60\% and 64\% less than in the present work, respectively) with a total of 6416 observed binary systems resulting also in a more populated matrix of 4.93\% observed entries. Therefore, the dataset used in this work allows for the analysis of the model in a chemical space of considerably higher diversity.  

The temperatures covered in the DECHEMA dataset range from -60 to 289.3 $^{\circ}$C. However, 90\% of the data points were measured between 20 and 120 $^{\circ}$C, and only 43.14\% of the binary systems in the data collection were measured at different temperatures within a range larger than 20 $^{\circ}$C. The overall distribution of $\ln \gamma_{ij}^\infty$ values ranges from -3.91 to 28.04. Among these, 22.28\% correspond to systems with negative deviations from ideality ($\gamma_{ij}^\infty < 1$) where the solvent-solute interactions are stronger than the solvent-solvent interactions, and 77.31\% correspond to positive deviations from ideality ($\gamma_{ij}^\infty > 1$) where the opposite is true. Only 0.41\% are reported as ideal systems potentially due to measurements that are within the experimental uncertainty range around $\gamma_{ij}^\infty = 1$. See sections S2 and S3 in the Supporting Information for the distribution of temperature and $\ln \gamma_{ij}^\infty$ values in the DECHEMA dataset, respectively. 

\subsection{Data splitting}
\label{sec:data_split}

In this work we used the method of stratified sampling for constructing the training and test sets. First, we classified all compounds using the chemical taxonomy of Classyfire \citep{djoumbou2016classyfire}. By doing this, the 1585 unique chemical compounds contained in the DECHEMA dataset were grouped into 91 chemical classes. Among them, the most common ones are ``Benzene and substituted derivatives" with 271 compounds and ``Organooxygen compounds" with 193 compounds. The complete list of chemical classes and the number of compounds contained in each class are available in section S4 of the Supporting Information. Using these 91 molecular classes a total of 841 binary combinations (e.g., solvent-class 1 with solute-class 32) were found. A random split (80/20) was performed on each one of these 841 bins of binary combinations to define the train and test sets. In case a bin contains a single solute-solvent pair this was placed on the training set. By using this stratified strategy we ensure that different types of molecular interactions are learned by the model and we enhance the analysis for its applicability domain by establishing specific chemical classes in which the model was actually tested.

\subsection{SMILES to molecular graph}
\label{smiles2graph}

Our method uses SMILES\citep{weininger1988smiles}, as string representation of molecules, to generate the corresponding solvent and solute graphs. In these graphs, the nodes and edges represent the atoms and chemical bonds, respectively. Initially, each node and edge is defined by a bit-vector of atomic $\mathbf{a} \in \{0, 1\}^{37}$ and bond $\mathbf{b} \in \{0, 1\}^{9}$ features  that are obtained using the cheminformatics package RDKit \citep{RDKit} (version 2021.03.1). These features are listed in Tables \ref{tbl:node_features} and \ref{tbl:edge_features} and are implemented as the concatenation of the one-hot-encoded vectors of each individual feature with the corresponding dimensions. In these vectors, the presence of the corresponding feature is indicated with the value 1 and the absence with the value 0. As a result, for each molecule the matrix of atoms' features $\mathbf{A} \in \{0, 1\}^{n_a \times 37}$ and the matrix of bonds' features $\mathbf{B} \in \{0, 1\}^{n_b \times 9}$ can be constructed, where $n_a$ and $n_b$ are the number of atoms and bonds in the molecule, respectively. Additionally, we specified the connectivity of the molecular graph via a matrix $\mathbf{C} \in \mathbb{N}^{2 \times 2n_b}$ of source and receiver node indexes. In this two-row matrix, the first row correspond to the indexes of the source nodes while the second row correspond to the indexes of the receiver nodes. Given that directed edges have no physical meaning on molecular graphs, source nodes act also as receiver nodes (cf. $2n_b$ on the dimensions of matrix $\mathbf{C}$). These 3 matrices are implemented as PyTorch tensors using the PyTorch Geometric library \citep{fey2019fast}. Hydrogen atoms are in general not included into the molecular graphs (rather they are explicitly included as atomic features, cf. ``Attached Hs" in Table \ref{tbl:node_features}), the exception to this is the molecular graph representing heavy water D$_2$O which includes the protium isotope of hydrogen.

\begin{table}
    \centering
  \caption{Atom features used to define the initial nodes in the molecular graphs. All features were implemented using one-hot-encoding.}
  \label{tbl:node_features}
  \begin{tabular}{lll}
    \hline
    Feature  & Description & Dimension  \\
    \hline
    Atom type & (C, N, O, Cl, S, F, Br, I, Si, Sn, Pb, Ge, H, P, Hg, Te) & 16 \\
    Ring & Is the atom in a ring? & 1 \\
    Aromatic & Is the atom part of an aromatic system? & 1 \\
    Hybridization & (s, sp, sp2, sp3) & 4 \\
    Bonds & Number of bonds the atom is involved in (0, 1, 2, 3, 4) & 5 \\
    Charge & Atom's formal charge (0, 1, -1) & 3 \\
    Attached Hs & Number of bonded hydrogen atoms (0, 1, 2, 3) & 4 \\
    Chirality & (Unspecified, clockwise, counter clockwise) & 3 \\
    \hline
  \end{tabular}
\end{table}

\begin{table}
\centering
  \caption{Bond features used to define the initial edges in the molecular graphs. All features were implemented using one-hot-encoding.}
  \label{tbl:edge_features}
  \begin{tabular}{lll}
    \hline
    Feature  & Description & Dimension  \\
    \hline
    Bond type & (Single, double, triple, aromatic) & 4 \\
    Conjugated & Whether the bond is conjugated & 1 \\
    Ring & Whether the bond is part of a ring & 1 \\
    Stereochemistry & (None, Z, E) & 3 \\
    \hline
  \end{tabular}
\end{table}

Previous works on GNNs for the prediction of $\gamma_{ij}^{\infty}$ have used similar atom and bond features \citep{D1DD00037C, rittig2022graph}. However, in a more general framework introduced by \cite{battaglia2018relational}, a graph can also include global-level features (i.e., features for the entire molecule). Therefore, under this scheme, a graph $G=(\mathbf{A}, \mathbf{B}, \mathbf{C}, \mathbf{u})$ is entirely defined by its node-features matrix $\mathbf{A}$, edge-features matrix $\mathbf{B}$, connectivity matrix $\mathbf{C}$ and global-features vector $\mathbf{u}$ (in our specific case $\mathbf{u} \in \mathbb{R}^3$). This idea of including global features explicitly can potentially support the modeling of molecular properties in mixtures (e.g., $\gamma_{ij}^\infty$) using GNNs. For example, as shown by \cite{qin2022capturing} and confirmed by the isothermal studies in the present work, by explicitly including hydrogen-bonding information into the learning framework better predictions for $\gamma_{ij}^\infty$ are achieved. Moreover, information regarding the complete molecular graphs (i.e., molecular descriptors) is nowadays easily accessible \citep{moriwaki2018mordred}. In this work, and by using this more general definition of molecular graphs, we have defined solvent and solute graphs that possess relevant global features and that are used to train in an end-to-end manner the proposed GH-GNN model for predicting $\gamma_{ij}^\infty$.

\begin{table}
\centering
  \caption{Global features used to define the initial global attributes of the molecular graphs.}
  \label{tbl:global_features}
  \begin{tabular}{lll}
    \hline
    Feature  & Description & Dimension  \\
    \hline
    Atomic polarizability & Polarizabilities for each atom & 1 \\
    Bond polarizability & Differences in atomic polarizabilities & 1 \\
    Topological polar surface area & 2D approximation of the polar surface area & 1 \\
    \hline
  \end{tabular}
\end{table}

Table \ref{tbl:global_features} shows the global features considered in this work. These features were inspired by the remarkable performance of the MOSCED model compared to other mechanistic and data-driven models in predicting $\gamma_{ij}^\infty$ \citep{brouwer2019model, D1DD00037C}. This performance can be attributed to the explicit inclusion of parameters strongly related to different types of molecular interactions: dispersion, induction, polarity, hydrogen-bonding acidity and hydrogen-bonding basicity. The dispersion parameters in MOSCED are mainly treated as regression parameters without direct physical meaning \citep{lazzaroni2005revision}. However, the induction, polarity and hydrogen-bonding parameters are related to their corresponding physical interactions as discuss by \citep{lazzaroni2005revision}. The induction parameter in MOSCED accounts for the ``dipole-induced dipole" or ``induced dipole-induced dipole" interactions that occur when compounds with pronounced polarizability are present in the liquid mixture. Inspired by this, the atomic and bond polarizability molecular descriptors (as calculated by the Mordred tool \citep{moriwaki2018mordred}) are included as part of the global attributes in the molecular graph. The atomic polarizability of the molecule is defined as the sum of the each individual atom's polarizability \citep{haynes2016crc} present in the molecule. Similarly, the bond polarizability of the molecule is defined as the sum of the absolute differences between the polarizability values of the pair of atoms present in each bond in the molecule. The polarity parameter in MOSCED is mainly related to the molecules' dipole moment and molar volume \citep{lazzaroni2005revision}. The dipole moment, specially, is dependent on the 3D conformation(s) of the molecule. However, given that conformer search calculations are still computationally expensive, a 2D approximation to the polar surface area \citep{prasanna2009topological} was included as global attribute via the topological polar surface area molecular descriptor (as calculated by the Mordred tool \citep{moriwaki2018mordred}). Hydrogen-bonding information is not included as global attribute to the graph, rather it is explicitly included as part of the edge features of the constructed ``mixture graph" which is explained in the next section.

\begin{figure}
\centering
  \includegraphics[width=12cm]{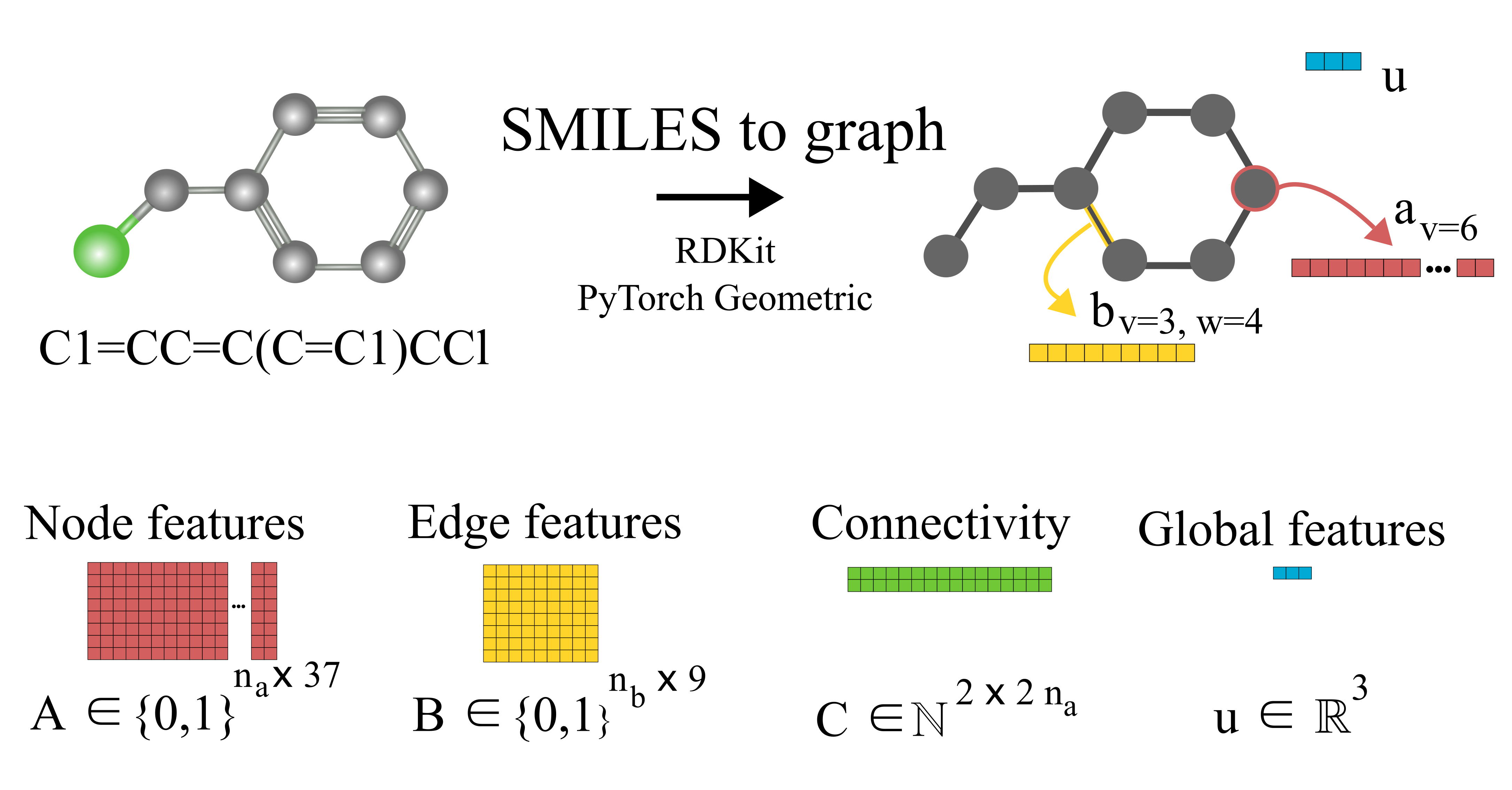}
  \caption{Schematic illustration of the molecular graph for benzyl chloride. First, the SMILES string is converted into an RDKit \citep{RDKit} molecular object where the atomic $a$ and bond $b$ features are obtained for each atom and bond in the molecule. In the proposed GH-GNN, global features $u$ are also included as part of the molecular graph \citep{battaglia2018relational}. These global features contain information of the polarizability and polarity properties of the molecule which are related to potential inductive and polar intermolecular interactions, respectively. Then, the node, edge and global features are collected and stored in tensors using PyTorch Geometric \citep{fey2019fast} to fully define the molecular graph. }
  \label{fgr:molecular_graphs}
\end{figure}

Figure \ref{fgr:molecular_graphs} shows an example of the SMILES to molecular graph procedure for benzyl chloride (C1=CC=C(C=C1)CCl). Without considering hydrogen atoms, benzyl chloride has 8 atoms (i.e., $n_a=8$) and 8 bonds (i.e., $n_b=8$). By using RDKit \citep{RDKit}, the atomic and bond features shown in Tables \ref{tbl:node_features} and \ref{tbl:edge_features} are calculated, one-hot-encoded and concatenated to construct the corresponding node $a_v$ and edge $b_{v,w}$ features vectors, respectively. In Figure \ref{fgr:molecular_graphs} the node-features vector of node 6 (i.e., $a_{v=6}$) is represented in red. Similarly, the features vector of the edge connecting node 3 and 4 (i.e., $b_{v=3, w=4}$) is represented in yellow. Notice that the specific numbering of the nodes is irrelevant since graphs do not have a defined order. The global features, as described in Table \ref{tbl:global_features}, of benzyl chloride are collected into vector $u$ represented here in blue. The node-features vector of each atom are stacked together to construct the matrix $A$, and similarly the edge features matrix $B$. The connectivity of the graph is stored in the matrix $C$, here represented in green. Therefore, starting with the SMILES string a set of three matrices $A$, $B$ and $C$ and one vector $u$ are constructed that define the molecular graph. 

\subsection{Mixture graph}

\begin{figure}
\centering
  \includegraphics[width=12cm]{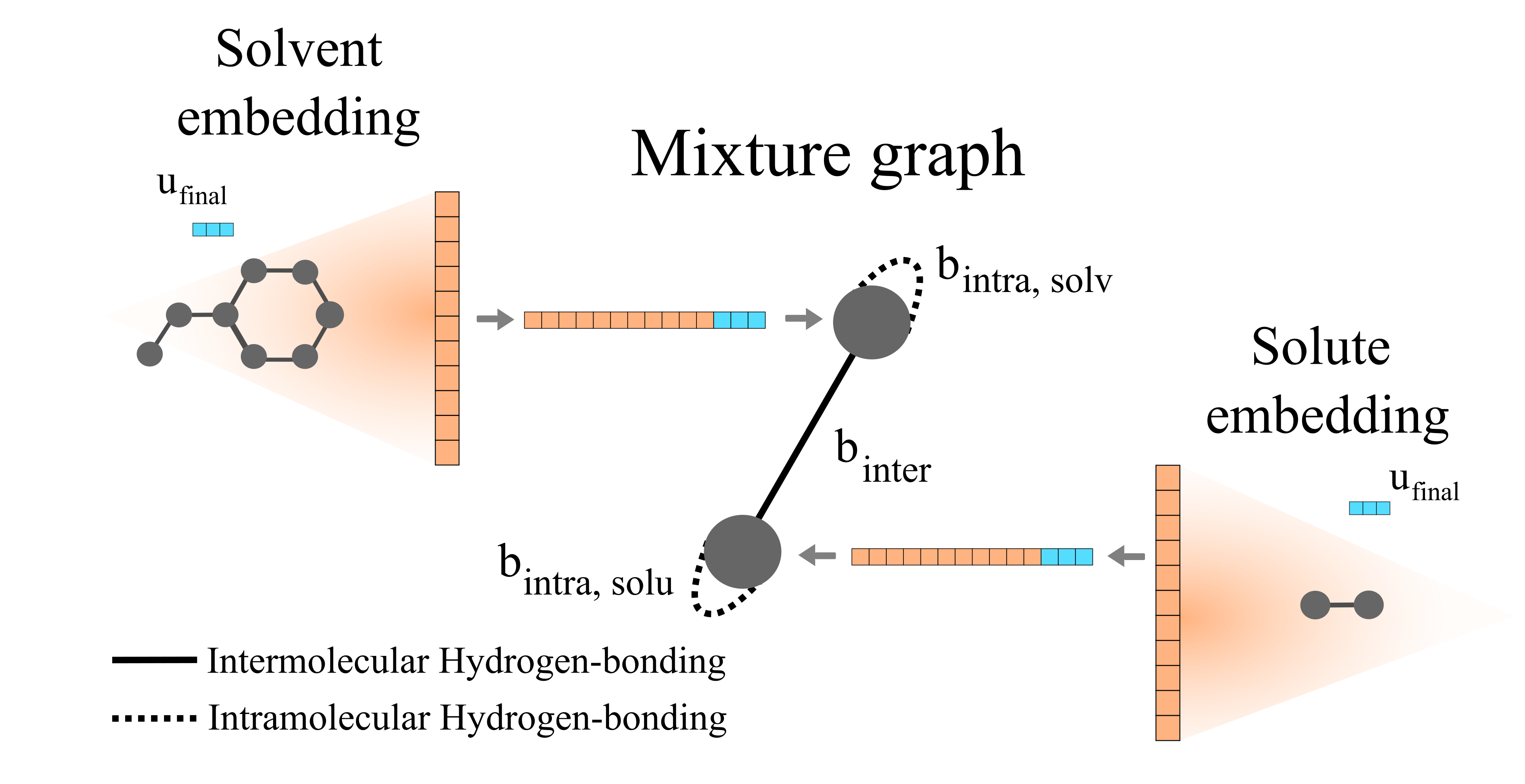}
  \caption{Schematic illustration of a two-component mixture graph. This graph has two nodes, one representing the solvent and the other representing the solute. The vector of node-features is defined by the final molecular fingerprint obtained after passing the molecular graph through a GNN and global pooling operations (here represented as the orange vector). In the case of using molecular graphs with global features $u$ (i.e., in our proposed GH-GNN), the concatenation of the global graph embedding (vector represented in orange) and the final global features embedding $u_{final}$ (represented in blue) is used as node-features for the mixture graph. The final global features embedding is obtained after passing the molecular graph through a GNN. As proposed by \cite{qin2022capturing}, the information of possible intramolecular interactions due to hydrogen-bonding is stored in the self-loops of the graph (represented as dotted edge lines). Similarly, this is done for intermolecular hydrogen-bonding interactions (represented as the solid edge line).}
  \label{fgr:mixture_graphs}
\end{figure}

 \cite{qin2022capturing} proposed the construction of a final graph (in this work refer to as the mixture graph) in which nodes represent chemical compounds that are present in the mixture and edges represent inter- and intramolecular interactions. The mixture graph is constructed using the learned embeddings of the solute and solvent graphs obtained after passing them through a GNN and a global pooling operation (see Figure \ref{fgr:mixture_graphs}). This concept allows for a more flexible learning scheme of molecular interactions using GNNs compared to our previously proposed \citep{D1DD00037C} simple concatenation of solute and solvent embeddings. As proposed by \cite{qin2022capturing}, we have included hydrogen-bonding information as edge features in the mixture graph. Information related to possible intermolecular hydrogen-bonding interactions is stored in the edge $\mathbf{b}_{inter}$ connecting the different nodes in the mixture graph (cf., solid edge line in Figure \ref{fgr:mixture_graphs}) and for a binary mixture is calculated by

\begin{equation}
\label{eqn:intermolecular_hb}
   \mathbf{b}_{inter} = 
   min 
   \left( 
       N^{HBA}_{solv}, N^{HBD}_{solu} 
   \right)
   + 
   min
   \left(
        N^{HBA}_{solu}, N^{HBD}_{solv}
   \right)
\end{equation}

\noindent where $N^{HBA}$ and $N^{HBD}$ stand for the number of hydrogen-bond acceptors and donors of the molecule, respectively. The subscripts $solv$ and $solu$ represent the solvent and solute species, respectively. By adding the minimum number of acceptors and donors between solvent and solute, the maximum number of hydrogen-bonding sites is captured. Similarly, the information related to possible intramolecular hydrogen-bonding interactions is stored in the self-loop edges $\mathbf{b}_{intra}$ (cf., dotted self-loop edge lines in Figure \ref{fgr:mixture_graphs}) and is calculated as

\begin{equation}
\label{eqn:intramolecular_hb}
   \mathbf{b}_{intra, k} = 
   min 
   \left( 
       N^{HBA}_{k}, N^{HBD}_{k} 
   \right)
\end{equation}

\noindent where $k$ stands for either the solvent or solute compound depending on which intramolecular interactions are being calculated.

Moreover, we have now extended this to the explicit inclusion of molecular descriptors related to the interactions due to dipole induction and polarity. These descriptors are first included as global features in the molecular graphs. After this, the graphs are passed through a GNN to obtain a graph with updated node, edge and global features (this step is explained in detail in the next section). Then, the global pooling embedding of the nodes of each graph (represented as orange vectors in Figure \ref{fgr:mixture_graphs}) is used as the corresponding node-features vector of the mixture graph. In case, global features are used (such in our proposed GH-GNN) the node-features vector of each node in the mixture graph is defined by the concatenation of the global pooling embedding of the nodes of the corresponding molecular graph and the final global feature embedding (cf. vector $u_{final}$ represented in blue in Figure \ref{fgr:mixture_graphs}).

\subsection{Graph Neural Networks}

GNNs perform a series of graph-to-graph transformations whose parameters can be optimized for a specific task (e.g., regression). Each of these transformations is usually referred to as a message passing layer. From layer to layer the number of nodes, edges and the connectivity of the graph stays the same. However, at each layer $l$, the vector embedding $\mathbf{a}_v$ representing the individual node $v$ is updated by combining its own information with the one of its neighbouring nodes $\mathcal{N}(v) = \{\mathbf{a}_w \vert ~ \mathbf{b}_{v,w} \in \mathbf{B}, v \neq w\}$ in a message passing scheme. In some more general implementations the connecting edges $\mathbf{b}_{v,w}$ are also updated from layer to layer \citep{battaglia2018relational}. In this way, after $l$ layers an updated graph is generated whose nodes now possess information regarding their $l$-level neighborhood. This message passing scheme is define by

\begin{equation}
\label{eqn:message_passing}
   \mathbf{a}^{(l+1)}_v = \mathcal{U}^{(l)} 
   \left( 
        \mathbf{a}^{(l)}_v, \underset{w \in \mathcal{N}(v)}{\mathcal{A}} M^{(l)}
            \left( 
            \mathbf{a}^{(l)}_v, \mathbf{a}^{(l)}_w,  f^E 
                \left( 
                \mathbf{b}^{(l)}_{v,w}
                \right)
            \right) 
    \right) 
\end{equation}

\noindent where $\mathcal{U}$ represents a differentiable function that updates the node embedding from layer $l$ to layer $l+1$; $\mathcal{A}$ stands for a differentiable and permutation invariant operator that aggregates all the messages coming from the neighbouring nodes $w \in \mathcal{N}(v)$; these messages are generated via a differentiable message function $M$; $f^E$ stands for a differentiable function that maps the edge embedding $\mathbf{b}_{v,w}$  to the same dimensions as the current node embeddings. Depending mainly on the specific definitions of the $\mathcal{U}$ and $M$ functions, many types of GNNs have been developed in the last years (e.g., GCN \citep{kipf2016semi}, Cheby \citep{defferrard2016convolutional}, MPNN \citep{gilmer2017neural}, GAT \citep{velickovic2017graph}, GIN \citep{xu2018powerful}).

After this, and in the context of molecular property prediction, a global pooling phase is applied to get the final prediction of interest \citep{gilmer2017neural}. This global pooling consists of a permutation-invariant operation (e.g., sum, max, min or mean) that condenses all the final graph information into a single vector that defines the molecular fingerprint. This fingerprint is then passed to a feed-forward neural network that performs the prediction. This general framework allows for an end-to-end learning from the initial graph(s) to the property of interest that can be trained using backpropagation. Due to their flexibility and overall good performance, many applications of GNNs to molecular property prediction have been investigated in recent years and reviewed in the literature \citep{wieder2020compact, yang2019analyzing}. All previous works on GNNs for predicting $\gamma_{ij}^\infty$ have mainly focused on this framework. 

\begin{figure}
  \includegraphics[width=16cm]{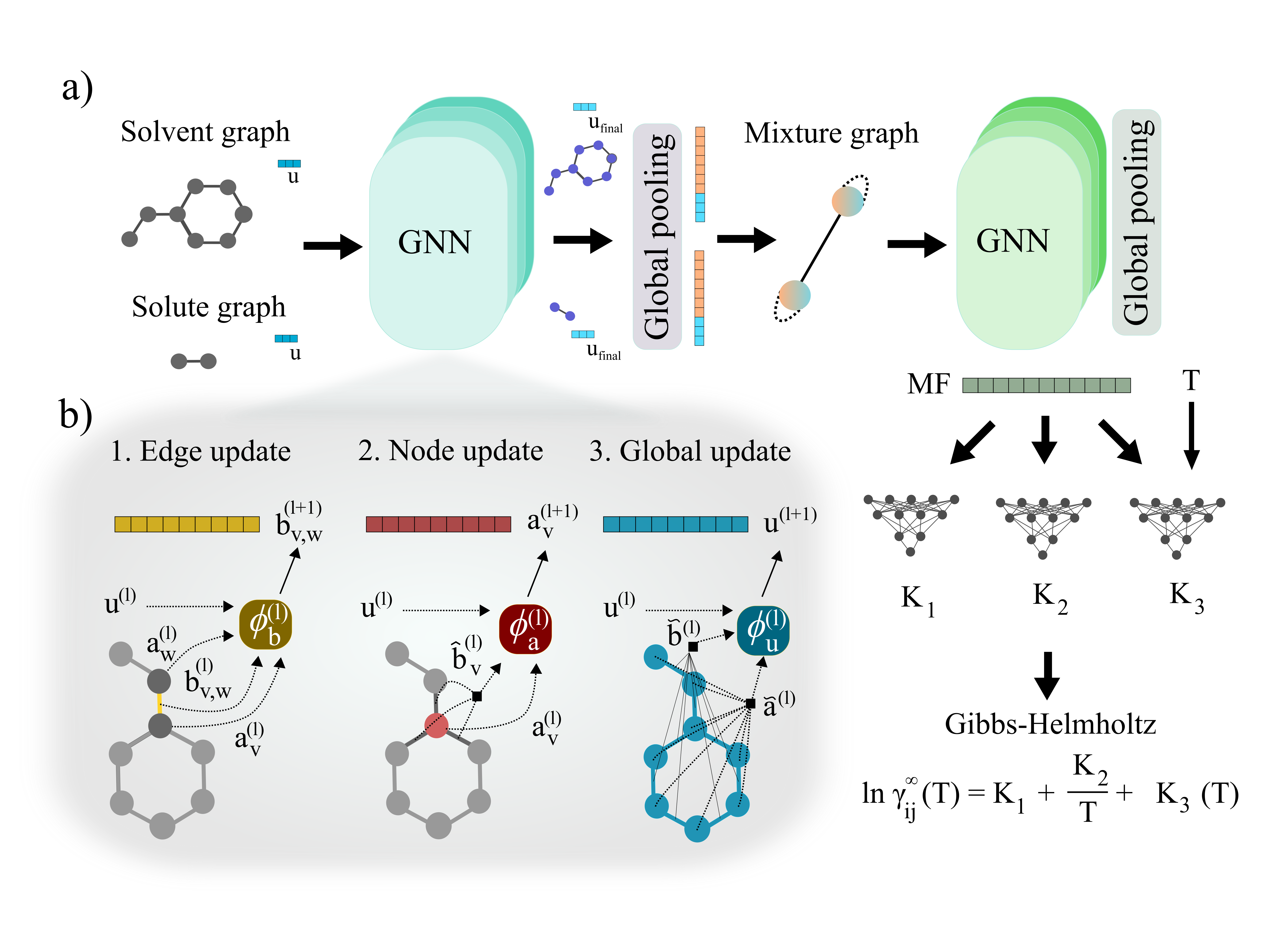}
  \caption{Schematic illustration of the Gibbs-Helmholtz Graph Neural Network (GH-GNN) model proposed in this work. a) First, the solute and solvent molecular graphs are passed through a GNN which updates their edge, node and global features. Then, the solute and solvent global embeddings are obtained by applying the global pooling operation. These embeddings are concatenated with the corresponding updated global features to define the node-features of the mixture graph. Hydrogen-bonding information is used to define the edge-features of the mixture graph. Afterwards, the mixture graph is passed through a GNN with an architecture originally proposed by \cite{gilmer2017neural} and used by \cite{qin2022capturing}. After that, a global pooling operation is performed to obtain a mixture fingerprint $MF$ that is used to calculate the $K_1$, $K_2$ and $K_3$ parameters of the integrated Gibbs-Helmholtz equation. b) Schematic illustration of the edge (Eq. \ref{eqn:edge_update}), node (Eq. \ref{eqn:node_update}) and global (Eq. \ref{eqn:global_update}) update operations carried out by the first GNN operating on the molecular graphs.}
  \label{fgr:graph_networks}
\end{figure}

However, in the more general framework introduced by \cite{battaglia2018relational}, GNNs can operate on graphs that also have global-level features. This framework of global-attributed graphs was employed for constructing the here proposed GH-GNN model. Here, all node, edge and global features are updated from layer to layer. Therefore, a graph $G^{(l)}=(\mathbf{A}^{(l)}, \mathbf{B}^{(l)}, \mathbf{C}, \mathbf{u}^{(l)})$ is updated to graph $G^{(l+1)}=(\mathbf{A}^{(l+1)}, \mathbf{B}^{(l+1)}, \mathbf{C}, \mathbf{u}^{(l+1)})$ using update functions for the edge (Eq. \ref{eqn:edge_update}), node (Eq. \ref{eqn:node_update}) and global (Eq. \ref{eqn:global_update}) attributes. First, each edge embedding $\mathbf{b}_{v,w}$ is updated using the embeddings of the connecting nodes $v$ and $w$, the current edge embedding itself, and the global-level embedding by

\begin{equation}
\label{eqn:edge_update}
   \mathbf{b}_{v,w}^{(l+1)} = \phi_b^{(l)}  
   \left( 
        \mathbf{a}_{v}^{(l)} \mathbin\Vert \mathbf{a}_{w}^{(l)} \mathbin\Vert \mathbf{b}_{v,w}^{(l)} \mathbin\Vert \mathbf{u}^{(l)}
   \right)
\end{equation}

\noindent where $\mathbin\Vert$ denotes concatenation of the embedding vectors and $\phi_b$ stands for (the edge update function) a single hidden layer neural network with the ReLU activation function. In Figure \ref{fgr:graph_networks}b a schematic representation of this edge update operation is shown for the yellow edge.  Second, the node embeddings $\mathbf{a}_{v}$ are updated using the updated attributes of the edges $\mathbf{b}_{v,w}^{(l+1)}$  on which the node $v$ is involved in, the current node embedding itself, and the global-level attributes by Eq. \ref{eqn:node_update}:

\begin{equation*}
   \widehat{\mathbf{b}}_{v}^{(l)} = \sum_{w \in \mathcal{N}(v) } \mathbf{b}_{v,w}^{(l+1)}
\end{equation*}

\begin{equation}
\label{eqn:node_update}
   \mathbf{a}_{v}^{(l+1)} = \phi_a^{(l)}  
   \left( 
        \mathbf{a}_{v}^{(l)} \mathbin\Vert \widehat{\mathbf{b}}_{v}^{(l)} \mathbin\Vert \mathbf{u}^{(l)}
   \right)
\end{equation}

\noindent where $\widehat{\mathbf{b}}_{v}$ stands for the sum of all updated edge embeddings that connect node $v$ with its neighbouring nodes $w \in \mathcal{N}(v)$ and $\phi_a$ stands for (the node update function) a single hidden layer neural network with the ReLU activation function. In Figure \ref{fgr:graph_networks}b a schematic representation of this node update operation is shown for the red node. Finally, the global embedding $\mathbf{u}$ is updated using its own previous information and the information of all updated nodes and edges in the molecular graph by by Eq. \ref{eqn:global_update}:

\begin{equation*}
   \widetilde{\mathbf{a}}^{(l)} = \frac{1}{n_a} \sum_{v=1}^{n_a} \mathbf{a}_{v}^{(l+1)}
\end{equation*}

\begin{equation*}
   \widetilde{\mathbf{b}}^{(l)} = \frac{1}{n_b} \sum_{k=1}^{n_b} \mathbf{b}_{k}^{(l+1)}
\end{equation*}

\begin{equation}
\label{eqn:global_update}
   \mathbf{u}^{(l+1)} = \phi_u^{(l)}  
   \left( 
        \mathbf{u}^{(l)} \mathbin\Vert \widetilde{\mathbf{a}}^{(l)}  \mathbin\Vert \widetilde{\mathbf{b}}^{(l)}
   \right)
\end{equation}

\noindent where $\widetilde{\mathbf{a}}$ and $\widetilde{\mathbf{b}}$ stand for the average pooling of all updated node and edge embeddings in the molecular graph, respectively; and $\phi_u$ stands for (the global update function) a single hidden layer neural network with the ReLU activation function. In Figure \ref{fgr:graph_networks}b a schematic representation of this global update operation is shown for the entire benzyl chloride graph in blue. The updating process of the global features naturally modifies their original physical interpretation (i.e., polarizability and topological surface area of the molecule), but it allows the GNN to learn relevant information that travels across the complete graph structure during the graph convolutions.

\subsection{Isothermal studies}

With the aim of comparing the performance of previous GNN architectures and mechanistic models for the prediction of $\ln \gamma_{ij}^\infty$, 9 isothermal studies were performed. For all these studies, the natural logarithm of $\gamma_{ij}^\infty$ was employed instead of the actual $\gamma_{ij}^\infty$ value for several reasons. First, it provides a better scaling of the data given the large range of values that $\gamma_{ij}^\infty$ can take depending on the species involved (e.g., when water is present significantly larger values are encountered compared to other compounds). Second, the logarithmic value appears naturally in thermodynamic equations when calculating chemical potentials. Third, when re-scaling the values by applying the exponential function, only positive values for $\gamma_{ij}^\infty$ are obtained. This meets the physical constraint of only having positive $\gamma_{ij}^\infty$ values. 

\begin{table}
\centering
  \caption{Information of the isothermal datasets from DECHEMA used for model comparison in the isothermal studies of this work.}
  \label{tbl:isothermal_info}
\begin{tabular}{cccccc}
\hline
\textbf{T ($^\circ$C)} &  \textbf{N$_\text{{solutes}}$} &  \textbf{N$_\text{{solvents}}$} &  \textbf{Size of matrix} & \textbf{N$_\text{{obs}}$} & \textbf{\% observations} \\
\hline
20 &          425 &           182 &           77,350 &    1,547  &      2.00  \\
25 &          638 &           419 &          267,322 &    3,716  &      1.39  \\
30 &          494 &           563 &          278,122 &    3,810  &      1.37  \\
40 &          497 &           343 &          170,471 &    2,455  &      1.44  \\
50 &          547 &           314 &          171,758 &    2,679  &      1.56  \\
60 &          527 &           367 &          193,409 &    2,766  &      1.43  \\
70 &          529 &           202 &          106,858 &    1,849  &      1.73  \\
80 &          553 &           209 &          115,577 &    1,745  &      1.51  \\
100 &         371 &           176 &           65,296 &    1,326  &      2.03  \\
\hline
\end{tabular}
\end{table}

A measurement was considered as isothermal at temperature T if it lays within the interval of T $\pm$ 0.5 $^\circ$C. With this definition, only 9 temperatures were found to have enough data (i.e., we define this as each temperature bin should contain at least 1000 data points). Table \ref{tbl:isothermal_info} summarizes the information of each isothermal dataset by providing the number of solutes and solvents present in the dataset, the size of the complete solute-solvent matrix ($N_{solutes} \times N_{solvents}$), the number of observations and the percentage of solute-solvent combinations that were actually measured and contained in the dataset. All the isothermal datasets were constructed out of the pre-splitted (train/test) DECHEMA dataset (see previous section on data splitting). For each one of the 9 different temperature levels, the points labeled as ``train" were used for training and the points label as ``test" were used for testing. The overall ratio of 80/20 was preserved in each of the 9 isothermal datasets.  

We compare the performance of the mechanistic models UNIFAC-Dortmund \citep{gmehling1998modified, lohmann2001unifac}, COSMO-RS \citep{eckert2002fast} and MOSCED \citep{lazzaroni2005revision}, to the performance of our previous GNN architecture \citep{D1DD00037C} (GNNprevious) and the SolvGNN \citep{qin2022capturing} architecture. A baseline random forest model was also trained on the concatenation of the pair of solute-solvent Morgan fingerprints \citep{morgan1965generation} with a radius of 4 and a size of 1024 bits for comparison. This simple baseline serves two main purposes: first, it shows whether or not the current mechanistic models can be beaten by a simple data-driven approach, and second, whether a more complex data-driven approach based on GNNs improves the prediction performance.  The hyperparameters for GNNprevious and SolvGNN were selected using Optuna \citep{optuna_2019} after 100 trials using 10-fold cross-validation on the training set. The specifications for the hyperparameter search and the final hyperparameters are available in Sections S5 and S6 of the Supporting Information for GNNprevious and SolvGNN, respectively. For the GNN-based models, we optimized the same hyperparameters that the previous implementations \citep{D1DD00037C, qin2022capturing} tuned and kept the rest of the hyperparameters with their original values. We used the Mean Squared Error (MSE) as the loss function during the training and the Adam optimizer \citep{kingma2014adam}. All the experiments were performed on a single NVIDIA Tesla P100 GPU (16 GB). 

The main goal of these isothermal studies is to assess the performance of GNN-based models that were recently proposed in the literature and to expand the discussion of these previous works \citep{D1DD00037C, qin2022capturing} to temperatures other than 25 $^\circ$C, and to compare them with common mechanistic models. However, these studies are limited to isothermal predictions. A more general framework that is able to predict $\ln \gamma_{ij}^\infty$ as a function of the temperature has still to be developed. For this, the following temperature dependency studies are carried out.

\subsection{Temperature dependency studies}

The temperature dependency of $\gamma_{ij}^\infty$ can be directly derived from the Gibbs-Helmholtz equation using the relation between the excess Gibbs free energy $g^E_{ij}$ and the activity coefficient (i.e., $g^E_{ij} = RT \ln \gamma_{ij}$)

\begin{equation*}
   \frac{\partial \left( g_{ij}^E / RT \right)}{\partial T} = - \frac{h_{ij}^{E, \infty}}{RT^2}
\end{equation*}

\begin{equation*}
   \frac{\partial \left( \ln \gamma_{ij}^\infty \right)}{\partial T} = - \frac{h_{ij}^{E, \infty}}{RT^2}
\end{equation*}

\begin{equation}
\label{eqn:gibbs_helmholtz}
   \frac{\partial \left( \ln \gamma_{ij}^\infty \right)}{\partial \left( 1/T \right)} = - \frac{h_{ij}^{E, \infty}}{R}
\end{equation}

\noindent where $h_{ij}^{E, \infty}$ stands for the molar excess enthalpy at infinite dilution. Under the assumption that the excess enthalpy is constant over the temperature range of interest, the above equation can be solved for $\ln \gamma_{ij}^\infty$ obtaining an explicit expression for the temperature dependency

\begin{equation}
\label{eqn:gibbs_helmholtz_constantH}
   \ln \gamma_{ij}^\infty (T) = K_1 + \frac{K_2}{T}
\end{equation}

\noindent where $K_1$ and $K_2$ are temperature independent constants for the specific solute-solvent system. In this expression, $K_1$ corresponds to the logarithmic activity coefficient at infinite dilution of the system when the temperature level approaches infinity, and $K_2=h^{E, \infty}_{ij}/R$, where $R$ is the universal gas constant. The assumption of constant excess enthalpy is often a good approximation for several systems of interest \citep{damay2021predicting, poling2001properties}. However, this assumption breaks when the range of studied temperatures is broader (e.g., in a range of around 40 $^\circ$C for some water-containing mixtures \citep{atik2004measurement, haidl2020activity}). This could potentially be circumvented by modeling the molar excess enthalpy at infinite dilution using dedicated data, which is left here as a future research direction. 

\subsubsection{Gibbs-Helmholtz Graph Neural Network (GH-GNN)}

In the proposed approach, we combine the physical knowledge of Eq. \ref{eqn:gibbs_helmholtz_constantH} with a set of Graph Neural Networks in a model that we call the Gibbs-Helmholtz Graph Neural Network (GH-GNN) for predicting $\ln \gamma_{ij}^\infty$ as a function of temperature. This framework is represented schematically in Figure \ref{fgr:graph_networks}a. First, the solvent and solute graphs are passed through a GNN that uses Equations \ref{eqn:edge_update}, \ref{eqn:node_update} and \ref{eqn:global_update} for updating the node, edge and global features of the corresponding graph, respectively. These updates are carried out over two message passing layers. After each message passing layer the updated graph is normalized using GraphNorm \citep{cai2021graphnorm}.  Afterwards, the final node features in the solute and solvent graphs are passed through a global mean pooling operation to obtained a vectorial representation of each graph. This vectorial representation of the graph is concatenated to the final global-features vector of the graph to define the final molecular fingerprint (represented as the half-blue-half-orange vectors in Figure \ref{fgr:graph_networks}a). The solute molecular fingerprint defines one of the nodes in the mixture graph, and the molecular fingerprint of the solvent does it for the other node in the mixture graph. The intermolecular hydrogen-bonding  information (Eq. \ref{eqn:intermolecular_hb}) defines the edge between the two nodes in the mixture graph, and each of the species' intramolecular hydrogen-bonding information (Eq. \ref{eqn:intramolecular_hb}) defines the corresponding self-loop edges. This mixture graph is passed through a GNN with the same architecture as originally proposed by \cite{gilmer2017neural}. This GNN is defined by Eq. \ref{eqn:message_passing} with $\mathcal{U}$ being a gated recurrent unit (GRU \citep{cho2014properties}) and $M=W^{(l)}a^{(l)}_v + \sum_{w \in \mathcal{N}(v)} a_w^{(l)} \phi_e (b_{v,w})$. Here $W$ is a learnable weight matrix and $\phi_e$ denotes a single-hidden layer neural network with the ReLU activation function that maps the edge-features' dimensions to the node-features' dimensions at the current layer $l$. The nodes of the updated final mixture graph (i.e., after passing the mixture graph through the second GNN) are passed through a global pooling operation that consist of the simple concatenation of both node embeddings to define the mixture fingerprint (represented as a green vector in Figure \ref{fgr:graph_networks}a). 

Finally, this mixture fingerprint is used to get the final prediction of $\ln \gamma_{ij}^\infty$ via predicting the parameters $K_1$ and $K_2$ of the Gibbs-Helmholtz-derived equation \ref{eqn:gibbs_helmholtz_constantH}. The assumption of a constant excess enthalpy is relaxed in our model by introducing a third parameter $K_3$ (cf. Fig. \ref{fgr:graph_networks}a) which is temperature-dependent, aiming to better capture the non-linearity of the temperature dependency of $\ln \gamma_{ij}^\infty$ (Eq. \ref{eqn:final_GNNGH_equation})

\begin{equation}
\label{eqn:final_GNNGH_equation}
   \ln \gamma_{ij}^\infty (T) = K_1 + \frac{K_2}{T} + K_3(T)
\end{equation}

\noindent the constants $K_1$, $K_2$ and $K_3$ are obtained after passing the mixture fingerprint through three separate multi-layer perceptrons (one for each constant). We used 2-hidden layer neural networks with the ReLU activation function. In the case of the neural network that computes $K_3$, the mixture fingerprint is multiplied by the temperature (measured in K) to introduce the temperature dependency to the parameter. The temperature in Equation \ref{eqn:final_GNNGH_equation} is used in K. To benchmark our GH-GNN model, we have used the SolvGNN architecture \citep{qin2022capturing} and we have extended it to capture the temperature dependency via Equation \ref{eqn:gibbs_helmholtz_constantH} and to include the global-level features as part of the nodes in the mixture graph (SolvGNNGH model). The hyperparameters of GH-GNN and SolvGNNGH were tuned using Optuna \citep{optuna_2019} over 100 trials using 10-fold cross-validation on the training set. The final hyperparameters and the ranges explored are available in Section S7 of the Supporting Information. We used the Mean Squared Error (MSE) as the loss function during the training and the AdamW optimizer \citep{loshchilov2017decoupled}. All these numerical studies were performed on a single NVIDIA Tesla P100 GPU (16 GB). 

In the matrix completion method (MCM) presented by \cite{damay2021predicting}, Eq. \ref{eqn:gibbs_helmholtz_constantH} has been also used to capture the temperature dependency of $\ln \gamma_{ij}^\infty$. However, compared to the MCM in which only systems which were measured over at least three different temperatures can be used for training, the GH-GNN model is able to use all the available data to simultaneously extract the temperature dependency and the relevant solute-solvent fingerprints. For instance, if a solute-solvent system was measured at a single temperature the MCM method has to discard it from the training, while the GH-GNN will still use this information to learn how to extract relevant molecular fingerprints over a larger space of molecular compounds. Then, the temperature dependency for that system would need to be extrapolated using the information obtained by other systems measured at different temperatures. Therefore, systems measured over a large range of temperatures will serve the GH-GNN to learn not only relevant structural information about the solute-solvent systems, but also to learn the temperature dependency of $\ln \gamma_{ij}^\infty$. Moreover, systems measured over a small range of temperatures can still be used by GH-GNN to enlarge its applicability to different chemical systems by learning additional solute-solvent structural information and interactions. In this way, all the scarce and thus valuable experimental data can be exploited.

The GH-GNN approach splits the problem of predicting $\ln \gamma_{ij}^\infty$ into three linked steps: 1) learning relevant molecular representations, 2) learning relevant mixture representations and 3) learning the property of interest. These consecutive steps are coupled and learned in an end-to-end fashion. However, the relevant intermediate representations can potentially be used for transfer learning for modeling properties of interest where the data is even more scarce than for activity coefficients. 

\section{Results and discussion}
\subsection{Isothermal studies}

Table \ref{tbl:isothermal_results} shows the comparison on the performance of the mechanistic models UNIFAC-Dortmund, COSMO-RS and MOSCED with the GNNprevious and SolvGNN models for the prediction of $\ln \gamma_{ij}^\infty$ in a series of isothermal studies based on the mean absolute error (MAE). A random forest baseline model trained on Morgan fingerprints is also provided. As discussed in previous sections, the UNIFAC-Dortmund and MOSCED models are limited in their applicability domain and not all the systems can be predicted by these two models. For UNIFAC-Dortmund, only systems involving molecules that can be correctly fragmented into its pre-established groups can be predicted. Not only the fragmentation scheme and the individual group parameters, but also all the involved binary-interaction parameters need to be available. In the case of MOSCED, the number of available compound-specific model parameters is very limited \citep{lazzaroni2005revision}. As a result, systems that involve substances without available MOSCED parameters are not possible to predict. For this reason two different comparisons are provided in Table \ref{tbl:isothermal_results} that show the models' performance on systems in the test set that can be predicted by UNIFAC-Dortmund and MOSCED. 

\begin{table}
\centering
  \caption{Comparison of the performance of UNIFAC-Dortmund (UNIFAC (Do)), MOSCED, COSMO-RS, GNNprevious \citep{D1DD00037C} and SolvGNN \citep{qin2022capturing} models on a set of isothermal studies. The GNN-based models were trained and tuned for each different isothermal set separately (additional information on hyperparameter tuning is available in sections S5 and S6 for GNNprevious and SolvGNN, respectively). A random forest (RF) model trained on solute-solvent Morgan fingerprints is shown as a baseline. For each model, the mean absolute error (MAE) is reported corresponding to the intersection of all models in the test set for the indicated model. The percentage of systems that represent the intersection of all feasible systems in the test set is also indicated as a coverage percentage (CP). }
  \label{tbl:isothermal_results}
\begin{tabular}{cccccccc}
\hline
\hline
\multicolumn{8}{c}{ \textbf{UNIFAC (Do) feasible systems in the test dataset}} \\
\hline
T & CP &\multicolumn{6}{c}{Mean absolute error (MAE)} \\
 ($^\circ$C) & (\%) & RF & UNIFAC (Do) & COSMO-RS & MOSCED & GNNprevious & SolvGNN \\
\hline
 20 & 96.25 & 0.64 &         1.08 &      0.52 &    - &         0.37 &     \textbf{0.32} \\
 25 & 86.73 & 0.62 &         1.23 &      0.54 &    - &         0.34 &     \textbf{0.31} \\
 30 & 73.14 & 0.43 &         0.48 &      0.39 &    - &         0.20 &     \textbf{0.19} \\
 40 & 77.53 & 0.51 &         0.57 &      0.40 &    - &         0.24 &     \textbf{0.22} \\
 50 & 77.01 & 0.40 &         0.35 &      0.33 &    - &         0.20 &     \textbf{0.18} \\
 60 & 80.90 & 0.42 &         0.43 &      0.36 &    - &         0.20 &     \textbf{0.19} \\
 70 & 87.77 & 0.48 &         0.40 &      0.38 &    - &         \textbf{0.22} &     0.25 \\
 80 & 85.60 & 0.43 &         0.40 &      0.38 &    - &         0.21 &     \textbf{0.20} \\
100 & 86.03 & 0.38 &         0.33 &      0.32 &    - &         0.20 &    \textbf{0.18} \\
\hline
\hline
\multicolumn{8}{c}{ \textbf{MOSCED feasible systems in the test dataset}} \\
\hline
T & CP &\multicolumn{6}{c}{Mean absolute error (MAE)} \\
 ($^\circ$C) & (\%) & RF & UNIFAC (Do) & COSMO-RS & MOSCED & GNNprevious & SolvGNN \\
\hline
20 & 53.44 & 0.60 &         0.47 &      0.38 &    0.31 &         0.33 &     \textbf{0.26} \\
 25 & 46.94 & 0.41 &         1.05 &      0.34 &    0.29 &         0.21 &     \textbf{0.19} \\
 30 & 21.77 & 0.54 &         0.38 &      0.34 &    0.25 &         0.24 &     \textbf{0.19} \\
 40 & 30.97 & 0.57 &         0.65 &      0.33 &    0.26 &         0.28 &     \textbf{0.22} \\
 50 & 31.23 & 0.45 &         0.37 &      0.30 &    0.17 &         0.18 &     \textbf{0.15} \\
 60 & 27.02 & 0.43 &         0.32 &      0.27 &    0.22 &         0.17 &     \textbf{0.16} \\
 70 & 22.07 & 0.58 &         0.44 &      0.30 &    0.26 &         \textbf{0.22} &     0.28 \\
 80 & 15.18 & 0.44 &         0.24 &      0.34 &    0.32 &         0.16 &     \textbf{0.13} \\
100 & 16.91 & 0.60 &         0.22 &      0.28 &    0.36 &         \textbf{0.21} &     0.22 \\
\hline
\end{tabular}
\end{table}

\begin{figure}
\centering
  \includegraphics[width=13cm]{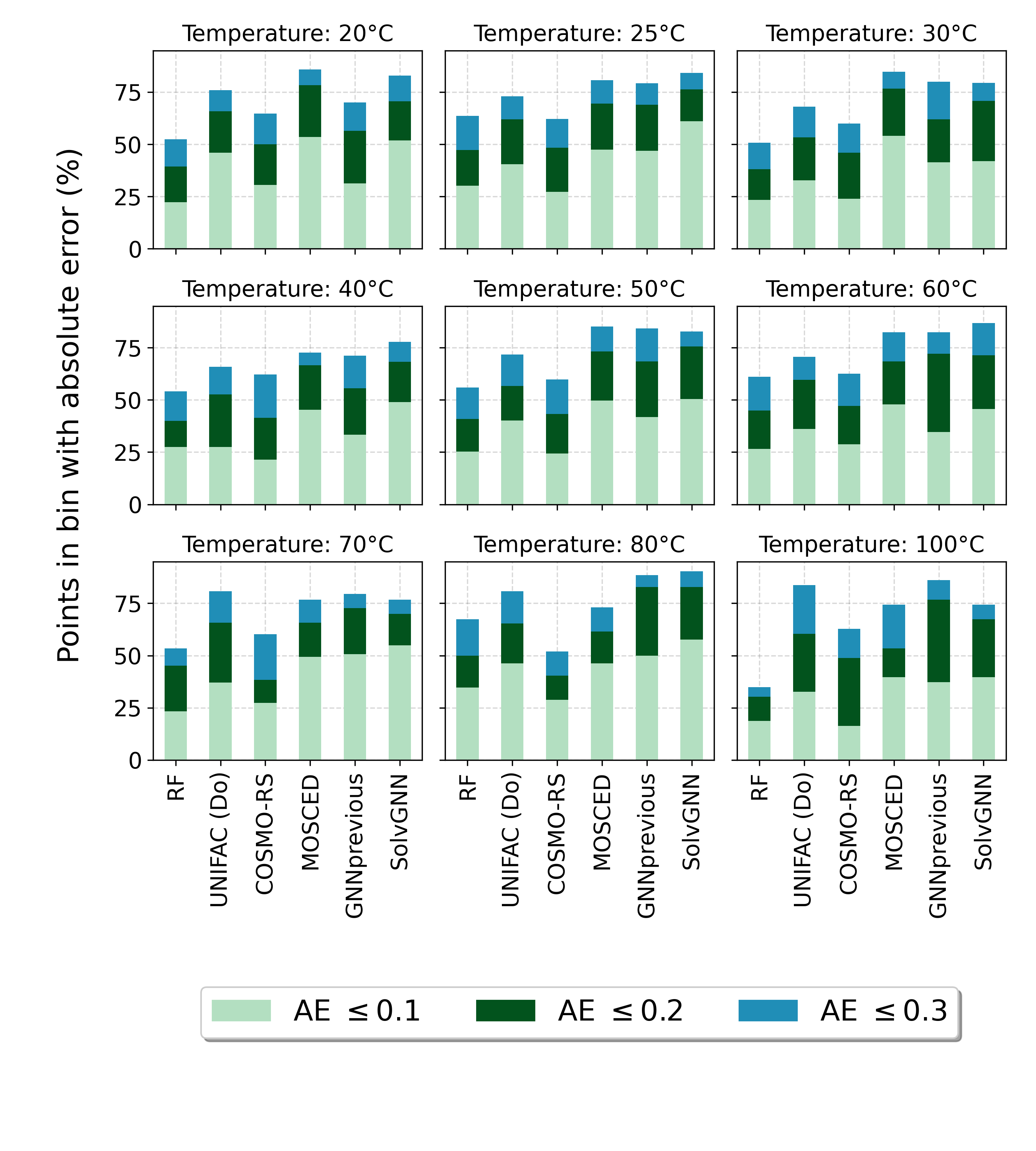}
  \caption{Percentage of systems in the test set that can be predicted by all models below the absolute error thresholds 0.1, 0.2 and 0.3 for each temperature. RF refers to the random forest model. AE stands for absolute error.}
  \label{fgr:isothermal_error_bins}
\end{figure}

Overall, the GNN-based methods achieve a lower MAE than the mechanistic methods. However, the comparison has to be carried-out with care also considering the differences on the type of systems that each model is able to predict. For instance, it should be noted that UNIFAC-Dortmund and COSMO-RS are able to also predict activity coefficients at finite dilution. For instance, when developing UNIFAC-Dortmund not only infinite dilution data was used, but also phase-equilibrium and caloric data were employed \citep{lohmann2001unifac}. Considering the feasible systems of MOSCED, it can be noted that MOSCED outperforms UNIFAC-Dortmund at lower temperatures. However, at high temperatures (i.e., 80 and 100 $^\circ$C) the accuracy of MOSCED starts to deteriorate. This can be explained by three main reasons. First, at 80 and 100 $^\circ$C the amount of feasible systems for MOSCED is considerably lower than at other temperatures which limits the robustness of the comparison. Second, the MOSCED parameters were regressed from a collection of experimental $\ln \gamma_{ij}^\infty$ values that include extrapolated values from phase-equilibrium measurements\citep{lazzaroni2005revision} which is known to produce poor estimations of the actual $\ln \gamma_{ij}^\infty$ values \citep{brouwer2021trends}. Third, the collection of experimental values used covers mainly lower temperatures (90\% of the data was measured between 20 and 82 $^\circ$C and only 6.85\% of the data was measured above 80 $^\circ$C) which suggests that the available MOSCED parameters are not trustworthy to extrapolate to high temperatures. It is also worth noting that SolvGNN achieves lower MAE compared to our previously proposed GNN architecture (GNNprevious) at almost all temperatures. This confirms, as pointed out by \cite{qin2022capturing}, that the explicit inclusion of relevant information about intermolecular interactions, in this specific case hydrogen-bonding interactions, is beneficial for modeling and predicting $\ln \gamma_{ij}^\infty$.

Figure \ref{fgr:isothermal_error_bins} reports the percentage of systems that are predicted below specific absolute error thresholds ($\leq0.1$, $\leq0.2$ and $\leq0.3$) for systems in the test set that can be predicted by all the five assessed methods. By looking at the percentage of systems that can be predicted with an absolute error below 0.1, MOSCED outperforms the rest of the models at $T=20^\circ$C, $T=30^\circ$C and $T=60^\circ$C, and performs similarly good as SolvGNN at $T=50^\circ$C and $T=100^\circ$C. It can be also observed, that SolvGNN outperforms the rest of the models at $T=25^\circ$C. This again highlights the advantage of including explicit information regarding molecular interactions (in MOSCED through model parameters related to induction, polarity and hydrogen-bonding interactions) into the model. The random forest baseline performs the worse among the models analyzed here suggesting that concatenating Morgan fingerprints does not capture the relevant interactions to accurately predict $\ln \gamma_{ij}^\infty$. The fact that the random forest model achieved lower MAE compared to UNIFAC-Dortmund at temperatures lower than 50 $^\circ$C (cf. Table \ref{tbl:isothermal_results}) can be explained by the the fact that UNIFAC-Dortmund severely mispredicts some of the systems which influence the MAE value (less robust metric to outliers).

One of the conclusions of our previous work \citep{D1DD00037C} regarding the better performance of GNN-based models for predicting $\ln \gamma_{ij}^\infty$ at 25 $^\circ$C compared to UNIFAC-Dortmund and COSMO-RS is here extended by analyzing a broad range of temperatures with practical interest. This can be seen in Figure \ref{fgr:isothermal_error_bins} by observing that UNIFAC-Dortmund and COSMO-RS are outperformed by at least one of the GNN-based models in the three error thresholds at each temperature.

\subsection{Temperature dependency studies}

Inspired by the observations of the isothermal studies in this work regarding the benefits of including explicit molecular interaction information into the modeling framework, we develop the Gibbs-Helmholtz Graph Neural Network (GH-GNN) for capturing the temperature dependency of $\ln \gamma_{ij}^\infty$. For benchmarking our proposed GH-GNN we develop an extension of SolvGNN \citep{qin2022capturing} that now incorporates the temperature dependency via Equation \ref{eqn:gibbs_helmholtz_constantH} in a model we call SolvGNNGH. GH-GNN has 2,547,765 trainable parameters and it lasts around 3.5 hours to be trained on a single P100 GPU, while SolvGNNGH has 843,203 trainable parameters and lasts around 2.7 hours to be trained on the same machine. The larger number of parameters of GH-GNN compared to SolvGNNGH is caused by the introduction of the edge and global updating functions and by the different hidden-layer sizes obtained from the hyperparameter tuning. 

Table \ref{tbl:temperature_results} presents the comparison results between GH-GNN and SolvGNNGH in terms of MAE and percentage of predicted systems with an absolute error lower than the thresholds 0.1, 0.2 and 0.3. The results are shown for the entire test dataset and for the systems in the test dataset that can be predicted by UNIFAC-Dortmund (around 84\% of the test set). This last comparison was made in order to benchmark both GNN-based models against this mechanistic model. The performance of other recently developed  data-driven models (i.e., matrix completion method (MCM)\citep{damay2021predicting} and SMILES-to-Property-Transformer (SPT) \citep{winter2022smile}) that predict $\ln \gamma_{ij}^\infty$ at varying temperatures is also presented. However, it has to be noted that the MCM and SPT models were trained and tested on different (and smaller) datasets. 

\begin{table}
\centering
  \caption{Comparison of the performance between GH-GNN and SolvGNNGH for the prediction of temperature-dependent $\ln \gamma_{ij}^\infty$. The results are reported for the test set and for the UNIFAC-Dortmund feasible systems in the test set based on the mean absolute error (MAE) and in the percentage of systems that are predicted with an absolute error below 0.1, 0.2 and 0.3 thresholds. In case of MAE, a lower value is better, and for the percentage error bins, higher is better. The best value in the comparison is marked as bold. The performance of other recently developed machine learning models (i.e., MCM and SPT) is shown at the bottom of the table. Their performance values are indicated in between parenthesis to highlight the fact that different data sets and methodologies were used to test those models and, hence, their results are not directly comparable. }
  \label{tbl:temperature_results}
\begin{tabular}{ccccc}
\hline
\hline
\multicolumn{5}{c}{ \textbf{Entire test dataset}} \\
\hline
Model &  MAE & $| \Delta \ln \gamma_{ij}^\infty | \leq 0.1$ & $| \Delta \ln \gamma_{ij}^\infty | \leq 0.2$ & $| \Delta \ln \gamma_{ij}^\infty | \leq 0.3$\\
\hline
SolvGNNGH  & 0.14 & 65.82\% & 83.63\% & 90.23\%  \\
GH-GNN      & \textbf{0.12} & \textbf{71.10\%} & \textbf{86.78\%} & \textbf{92.10\%}  \\
\hline
\hline
\multicolumn{5}{c}{ \textbf{UNIFAC (Do) feasible systems in the test dataset}} \\
\hline
Model & MAE & $| \Delta \ln \gamma_{ij}^\infty | \leq 0.1$ & $| \Delta \ln \gamma_{ij}^\infty | \leq 0.2$ & $| \Delta \ln \gamma_{ij}^\infty | \leq 0.3$\\
\hline
SolvGNNGH   & 0.14  & 65.15\%  & 83.2\%    & 89.89\%  \\
GH-GNN       & \textbf{0.12}  & \textbf{69.98\%}  & \textbf{86.19\%}   & \textbf{91.75\%}  \\
UNIFAC (Do) & 0.60  & 33.1\%   &  51.76\%  & 64.32\%  \\
\hline
\hline
\multicolumn{5}{c}{ \textbf{Other models in other datasets}} \\
\hline
MCM\citep{damay2021predicting}   & -  & -  & -    & (76.6\%)  \\
SPT\citep{winter2022smile}       & (0.11)  & -  & -   & (94\%)  \\
\hline
\end{tabular}
\end{table}

It can be seen that GH-GNN outperforms the rest of the models tested in the same dataset in all metrics. Considering the experimental $\ln \gamma_{ij}^\infty$ uncertainty estimation of \cite{damay2021predicting} of 0.1-0.2, the GH-GNN is able to predict more than 85\% of the systems within the experimental uncertainty. However, it is important to note that this is just an empirical estimation. A much more detailed and statistically significant analysis on the experimental uncertainty of the data is still necessary and left as a future research subject. The relatively high MAE achieved by UNIFAC-Dortmund is mainly caused by some severe miss-predictions (outliers) for systems containing pyridines or quinolines and their derivatives, and systems containing water. However, its performance is still poorer than the GNN-based models considering the percentage of systems predicted within the error thresholds (a metric more robust to outliers). It has to be pointed out that UNIFAC-Dortmund was trained on most of the data included in the test set of this work \citep{lohmann2001unifac}. Despite this fact, GH-GNN predicts $\ln \gamma_{ij}^\infty$ much more accurately. 

Moreover, it is worth noting that the availability of more experimental data increases the performance of the GNN-based models considerably. This is highlighted by the larger performance gap between UNIFAC-Dortmund and GNN-based models in the temperature dependency studies compared to the ones presented in the isothermal studies. Furthermore, the better performance of GH-GNN compared to SolvGNNGH suggests that besides the inclusion of hydrogen-bonding information, the inclusion of polarizability and polarity terms as global-level graph descriptors enhances the prediction quality of $\ln \gamma_{ij}^\infty$. The inclusion of such descriptors is potentially related to the ability of the model to capture "dipole-induced dipole" and "induced dipole - induced dipole" interactions in a similar way as the MOSCED model does it by using induction and polarity parameters \citep{lazzaroni2005revision}.   

The performance of SPT is roughly comparable to the one that GH-GNN shows. However, GH-GNN achieves this performance on a much larger and diverse chemical space (cf. Data cleaning section). Moreover, the SPT model is based on Natural Language Processing techniques that require a computationally very expensive pre-training step (using millions of data points simulated from COSMO-RS) and a fine-tuning step using experimental data. By contrast, GH-GNN is directly trained on experimental data achieving a similar performance interpolating within the chemical space covered during training. The accuracy of GH-GNN is likely to increase if a similar transfer learning pre-step is used.

\subsection{Continuous inter-/extrapolation}

In order to analyze the performance of GNN-based models on extrapolating to system's temperatures that are outside of the range of temperatures seen during the training phase for those specific systems, we have measured the error for three different tasks. First, the continuous interpolation of specific systems was studied. Here, all solute-solvent combination that was present in the training set within a range of temperatures from $T_1$ to $T_2$ and that was present in the test set with temperature(s) $T_{inter}$ such that $T_1 < T_{inter} < T_2$ was analyzed. The MAE on these interpolation systems on the test set is presented in Table \ref{tbl:continous_polation} for GH-GNN and SolvGNNGH. Second, the MAE is presented in Table \ref{tbl:continous_polation} for solute-solvent systems contained in the test set that were measured at temperature(s) $T_{extra, L}$ such that $T_{extra, L} < T_1 < T_2$ which are referred to as ``lower extrapolation" systems. Third, the extrapolation performance to systems in the test set measured at $T_{extra, U}$ such that $T_1 < T_2 < T_{extra, U}$ are referred to as ``upper extrapolation" systems in Table \ref{tbl:continous_polation}.

\begin{table}
\centering
  \caption{Performance of GNN-based models on the continuous inter-/extrapolation to different temperatures. Given a system in the training set that was measured between temperatures $T_1$ and $T_2$, interpolation is defined for the specific system in the test set measured at $T_{inter} \vert T_1 < T_{inter} < T_2$. Similarly, extrapolation to lower temperatures is defined for systems in the test set that were measured at $T_{extra, L} \vert ~ T_{extra, L} < T_1 < T_2$, and extrapolation to higher temperatures is defined for systems measured at $T_{extra, U} \vert ~ T_1 < T_2 < T_{extra, U}$. The results are reported for the mean absolute error (MAE), lower is better. The best value on the comparison is marked in bold. The number of data points for the corresponding inter/extrapolation sets are also shown along with the percentage that they represent of the entire test set.
  }
  \label{tbl:continous_polation}
\begin{tabular}{cccc}
\hline
\hline
\multicolumn{4}{c}{ \textbf{Interpolation on the test dataset ($T_1 < T_{inter} < T_2$)}} \\
\hline
Model &  MAE & Number of data points & Percentage of complete test set \\
\hline
SolvGNNGH  & 0.12  & 3025 & 36.41\% \\
GH-GNN      & \textbf{0.11}  & 3025 & 36.41\% \\
\hline
\hline
\multicolumn{4}{c}{ \textbf{Lower extrapolation systems on the test dataset ($T_{extra, L} < T_1 < T_2$)}} \\
\hline
Model &  MAE & Number of data points & Percentage of complete test set\\
\hline
SolvGNNGH  & 0.14  & 1954 & 23.52\% \\
GH-GNN      & \textbf{0.12}  & 1954 & 23.52\% \\
\hline
\hline
\multicolumn{4}{c}{ \textbf{Upper extrapolation systems on the test dataset ($T_1 < T_2 < T_{extra, U}$)}} \\
\hline
Model &  MAE & Number of data points & Percentage of complete test set\\
\hline
SolvGNNGH  & 0.12  & 1684 & 20.27\% \\
GH-GNN      & \textbf{0.11}  & 1684 & 20.27\% \\
\hline
\end{tabular}
\end{table}

GH-GNN performs better than SolvGNNGH for both interpolation and extrapolation to different temperatures. The performance on continuous inter-/extrapolation of these models is remarkable. This is a result of introducing the explicit temperature dependency into the GNN framework via the Gibbs-Helmholtz-derived equation. This equation provides a simple physics-based relationship that can be exploited in practice for introducing the temperature dependency of $\ln \gamma_{ij}^\infty$ to the model. This is in contrast to the approach of other works that introduce the temperature dependency via simple concatenation \citep{rittig2022graph} or by a projection of the temperature value to an embedding state within the model \citep{winter2022smile}.

\begin{figure}
  \includegraphics[width=17cm]{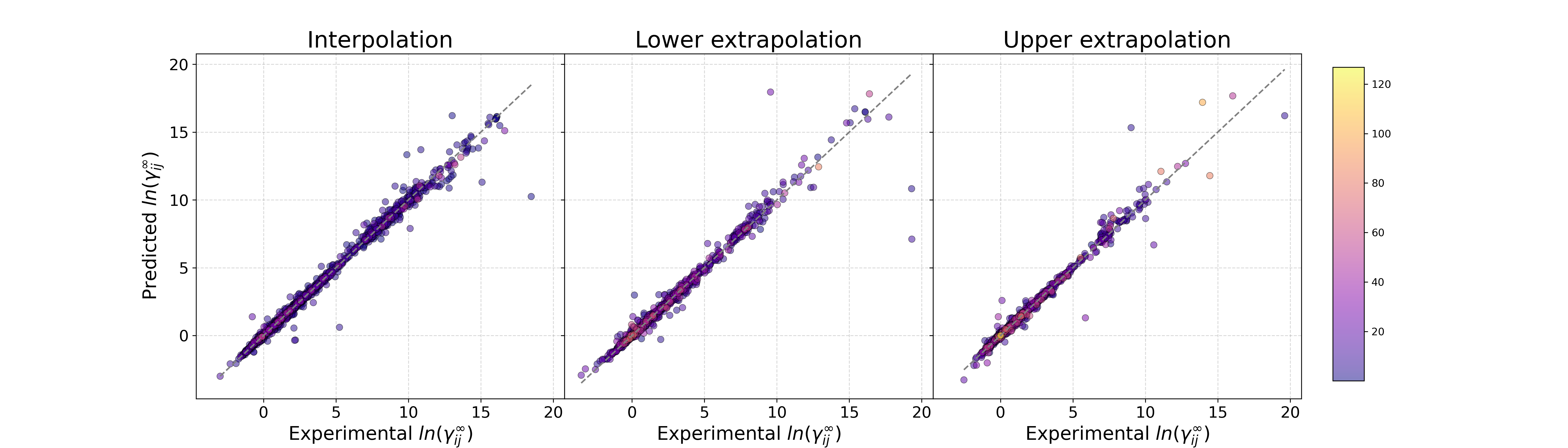}
  \caption{Parity plot of experimental vs. predicted $\ln \gamma_{ij}^\infty$ with the proposed GH-GNN model for continuous (left) interpolation systems in the test set (3025 data points), (center) extrapolation to lower temperatures (1954 data points) and (right) extrapolation to higher temperatures (1684 data points). The gray parity line corresponds to the perfect prediction. The color bar indicates the absolute distance between the interpolated temperature and the closest temperature in the training set.}
  \label{fgr:parity_T_complete}
\end{figure}

Figure \ref{fgr:parity_T_complete} shows the parity predictions for interpolation, lower extrapolation and upper extrapolation to other temperatures. The color map indicates the absolute distance of the inter-/extrapolated temperature to the closest temperature in the training set. It can be seen that the inter-/extrapolation distance is not correlated with the degree of accuracy in the prediction. For instance, in Figure \ref{fgr:parity_T_complete} many systems inter-/extrapolated to temperatures away from the training values by more than 60$^\circ$C are predicted with high accuracy. By contrast, in the three cases shown in Figure \ref{fgr:parity_T_complete} there exist systems with closer observations in the dataset that are less accurately predicted. This is likely an indication that the prediction performance of GH-GNN is mostly related to the complexity of the chemical compounds involved rather to the temperature conditions at which a given system is predicted. It is also noticeable that the predictions of the GH-GNN model start to deteriorate mainly on systems with $\ln \gamma_{ij}^\infty > 10$, which in their vast majority include water as solvent. This difficulty in predicting water-containing systems has been reported for decades \citep{kojima1997measuring} and it is still present when using modern models based on machine learning methods \citep{winter2022smile}. It has to be noted that data for $\ln \gamma_{ij}^\infty > 10$ systems is additionally a minor proportion compared to all the experimental data available (cf. section S3 in the Supporting Information) and it might be the main reason behind the poor prediction accuracy for this type of systems.    

\subsection{Discrete inter-/extrapolation}

As discussed in the previous section, the performance of GH-GNN is mainly dependent on the type of chemical species predicted rather than on the temperature conditions at which the prediction is made. For this reason, in this section we analyze the performance of GH-GNN on discrete interpolation and extrapolation i.e., with respect to different chemical compounds. 

\begin{figure}
     \centering
     \begin{subfigure}[b]{0.49\textwidth}
         \centering
         \includegraphics[width=\textwidth]{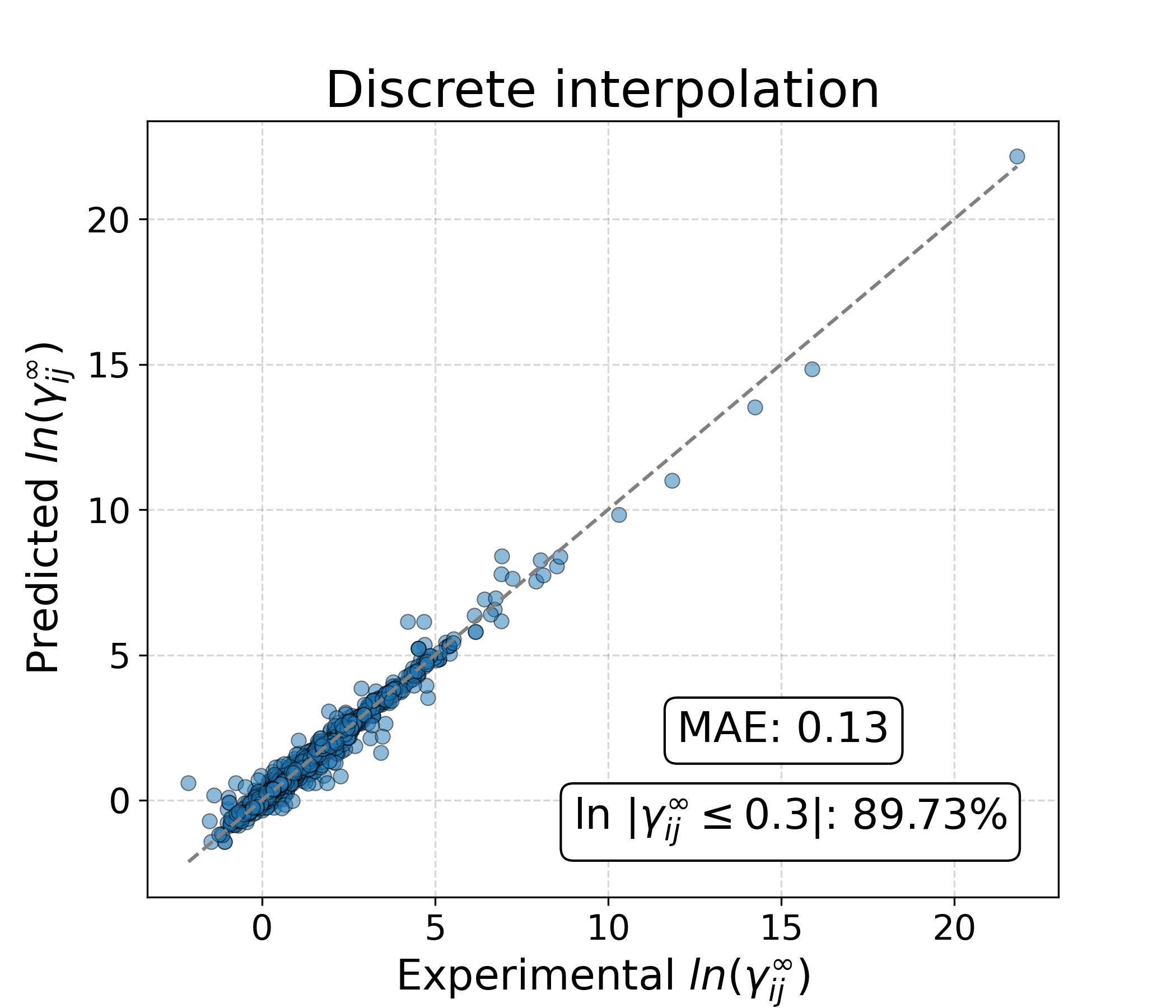}
     \end{subfigure}
     \hfill
     \begin{subfigure}[b]{0.49\textwidth}
         \centering
         \includegraphics[width=\textwidth]{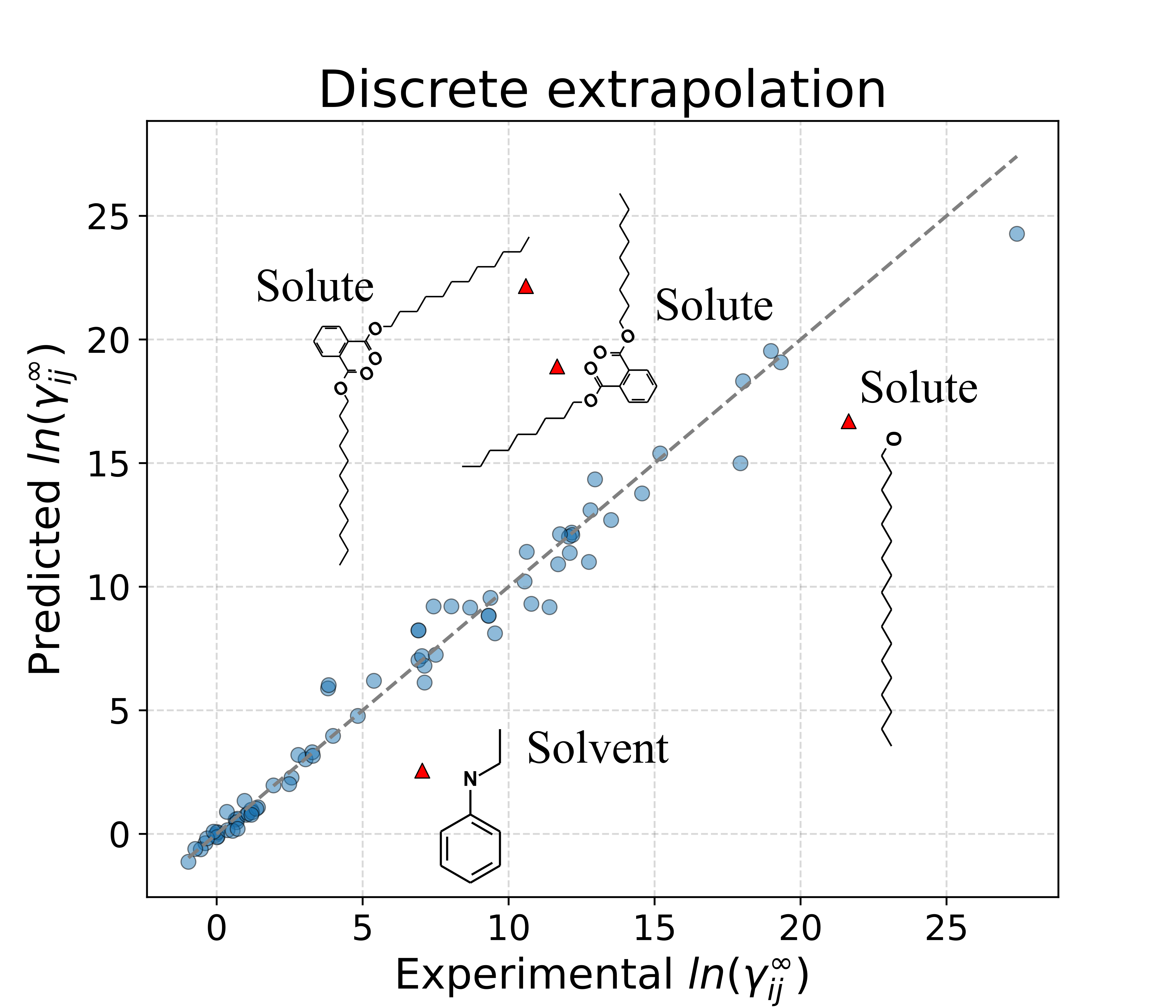}
     \end{subfigure}
    \caption{(Left) Parity plot of the experimental vs. predicted $\ln \gamma_{ij}^\infty$ with the proposed GH-GNN model for compound-related (discrete) interpolation systems (solute-solvent systems that are not present in the training set in this precise combination, but from which the solute and solvent are present in a different paring). This task corresponds to filling the missing entries of the solute-solvent training matrix of the matrix completion method (MCM) \citep{damay2021predicting}. (Right) Parity plot of the experimental vs. predicted $\ln \gamma_{ij}^\infty$ with the proposed GH-GNN model for compound-related (discrete) extrapolation systems in the original test set. The red triangles indicate the worst four predicted systems, all of them involving water and the shown compound as either solvent or solute as indicated in the corresponding labels. In both plots, the gray parity line corresponds to the perfect prediction.}
    \label{fgr:parity_discrete_inter_extra}
\end{figure}

For testing the discrete interpolation capabilities of GH-GNN we have analyzed all the solute-solvent systems in the test set that are not contained in these precise combinations within the training set, but the individual solute and solvent species are present in the training set in other pairings. As discussed by \cite{winter2022smile}, this interpolation task is comparable to the process of completing the solute-solvent matrix, as proposed in the last years \citep{damay2021predicting}. For illustration, Figure \ref{fgr:parity_discrete_inter_extra} (left) shows the parity plot for the GH-GNN predictions for compound-related (discrete) interpolation systems. The number of this type of binary-systems in the test set is 1568 for which the GH-GNN achieves a MAE of 0.13 with 89.73\% of the systems being predicted with an absolute error below 0.3. We can therefore conclude that GH-GNN is able to interpolate (i.e., predict the missing entries in the training solute-solvent matrix) with remarkable precision. For comparison, the MCM operated on a much smaller solute-solvent matrix predicts 76.6\% of the systems below an absolute error of 0.3 \citep{damay2021predicting}.

For analyzing the extrapolation capabilities of our proposed GH-GNN framework we considered the systems in the test set which are formed by either a solvent or a solute that is not present at all in the training set. In the original test set only 77 of such systems exist. Figure \ref{fgr:parity_discrete_inter_extra} (right) shows the parity plot of the GH-GNN predictions. In this figure, the worst four predicted systems are highlighted as red triangles. These four systems contain water as solvent, except for the prediction with the lowest $\ln \gamma_{ij}^\infty$ value which also contains it, but as a solute. For this last system, only two points in the training set contain the same combination of solute-solvent chemical classes and the Tanimoto similarity between the extrapolated solvent and any molecule in the training set was found to be 0.  Also, as depicted in Figure \ref{fgr:parity_discrete_inter_extra} (right) these systems contained large molecules as solutes which might explain the difficulty that GH-GNN has on predicting them. Again, the prediction accuracy diminishes for systems with $\ln \gamma_{ij}^\infty > 10$. GH-GNN achieves an overall MAE of 0.57 on these systems without the worst 4 predictions (red triangles in Figure \ref{fgr:parity_discrete_inter_extra} (right)). However, the number of extrapolation systems in the original test set is limited and, as a result, the conclusions for extrapolation cannot be confidently generalized. For this reason, we have gathered a new subset of the data originally collected from \cite{brouwer2021trends} and reviewed by \cite{winter2022smile} corresponding to the systems in which either the solute or the solvent is not present in our training set. This new dataset (referred to as ``external dataset") consists of experimentally measured $\ln \gamma_{ij}^\infty$ values, collected from the open-literature including only molecular systems (i.e., no ionic liquids). 

\begin{figure}
     \centering
     \begin{subfigure}[b]{0.49\textwidth}
         \centering
         \includegraphics[width=\textwidth]{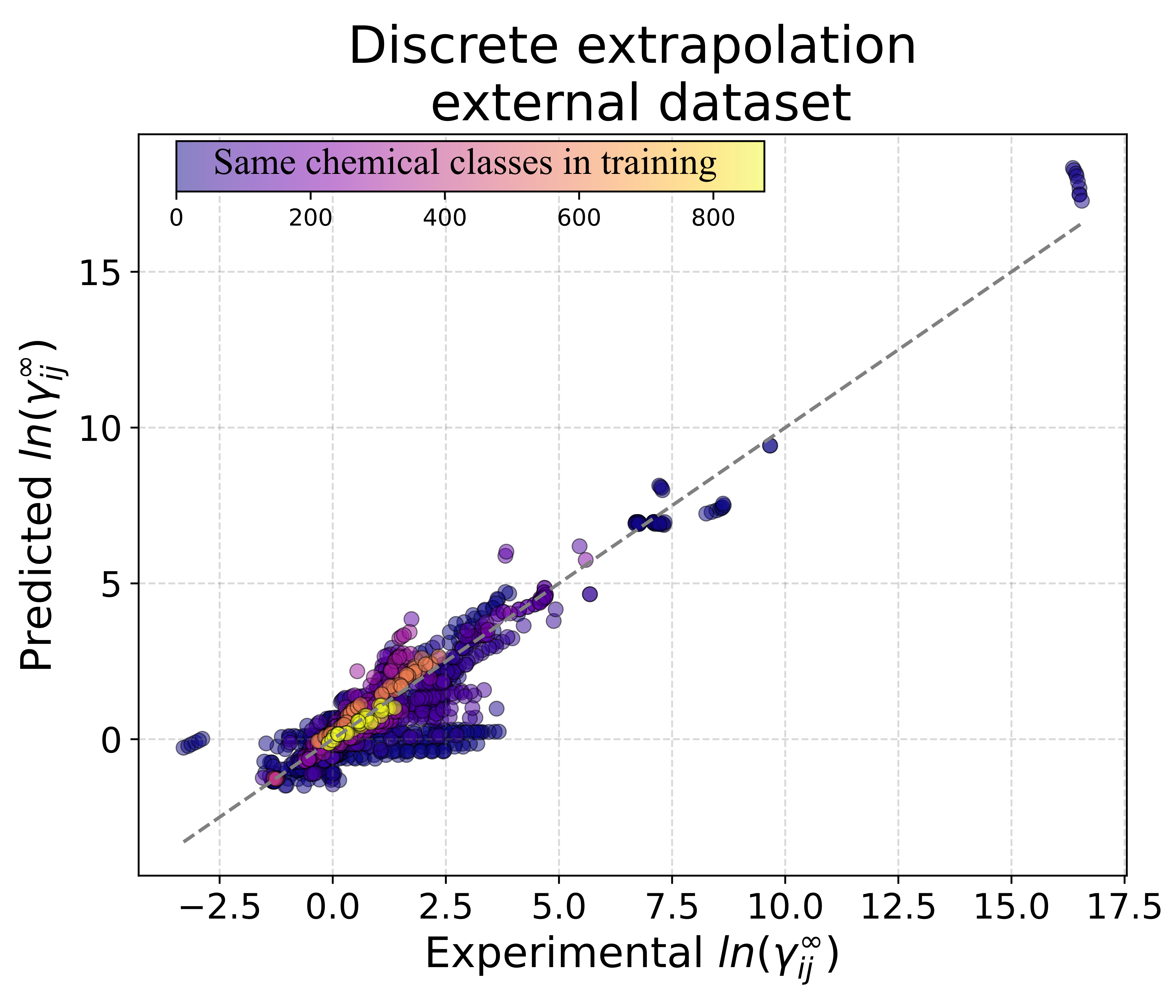}
     \end{subfigure}
     \hfill
     \begin{subfigure}[b]{0.49\textwidth}
         \centering
         \includegraphics[width=\textwidth]{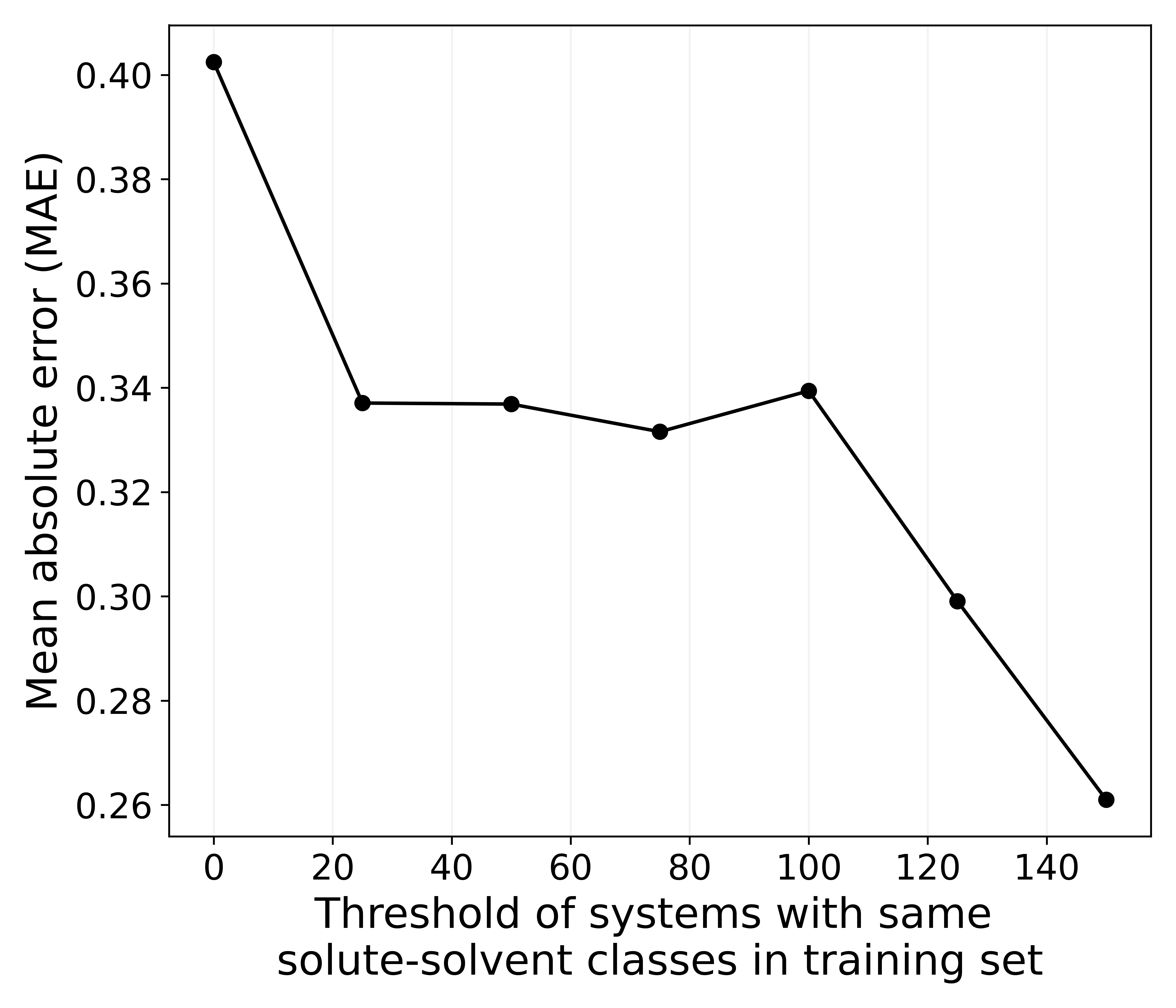}
     \end{subfigure}
    \caption{(Left) Parity plot of the experimental vs. predicted $\ln \gamma_{ij}^\infty$ with the proposed GH-GNN model for (discrete) extrapolation systems with respect to chemical compounds in the external dataset. The gray parity line corresponds to the perfect prediction. The color bar indicates the number of binary-systems in the training set with the same solute-solvent chemical classes according to the classification obtained from Classyfire \citep{djoumbou2016classyfire}. (Right) Mean absolute error (MAE) achieved by GH-GNN for systems in the external dataset for which the indicated minimum threshold of systems with the same solute-solvent classes are contained in the training set.}
    \label{fgr:parity_discrete_extrapolation_Brouwer_Nsystems_MAE}
\end{figure}

\begin{figure}
     \centering
     \begin{subfigure}[b]{0.49\textwidth}
         \centering
         \includegraphics[width=\textwidth]{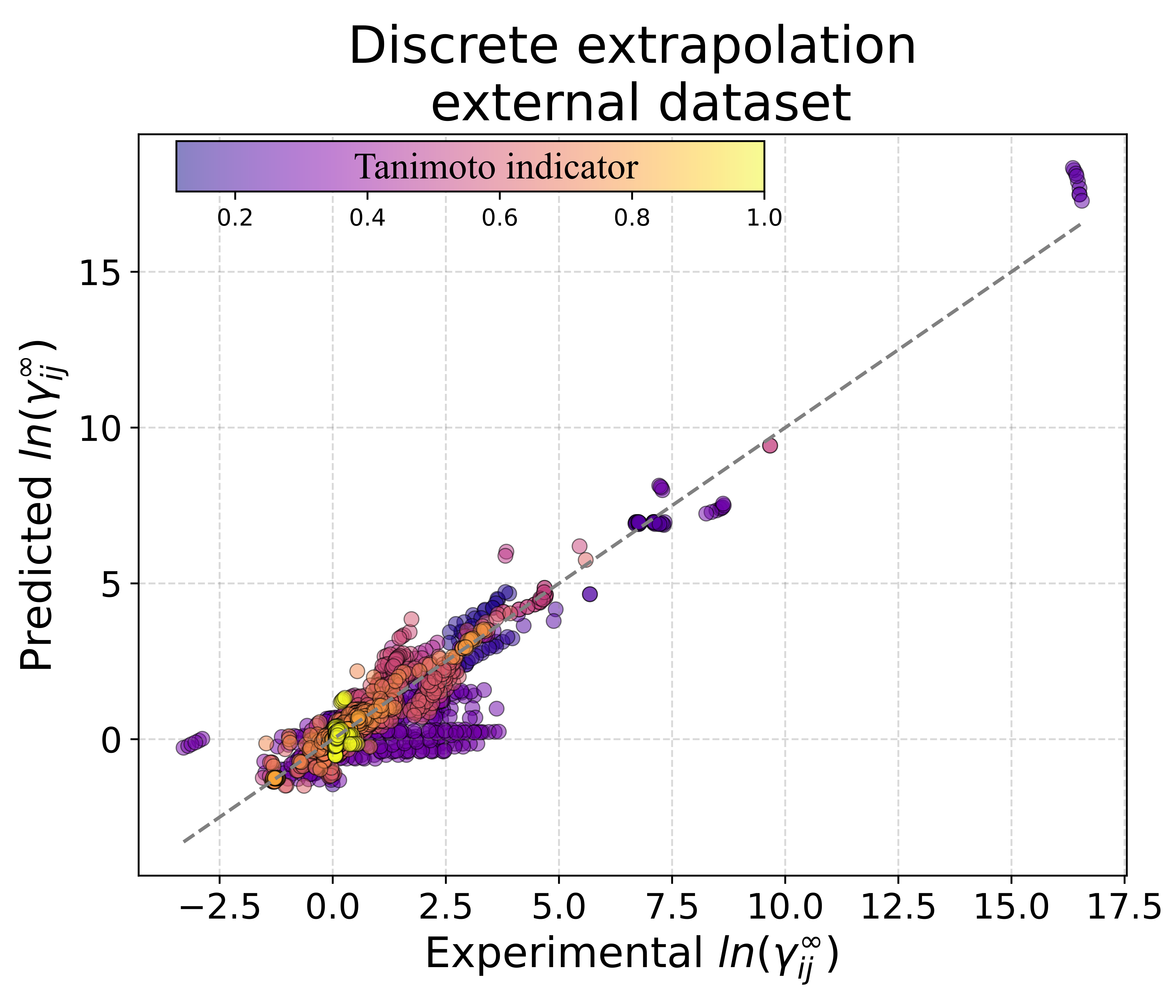}
     \end{subfigure}
     \hfill
     \begin{subfigure}[b]{0.49\textwidth}
         \centering
         \includegraphics[width=\textwidth]{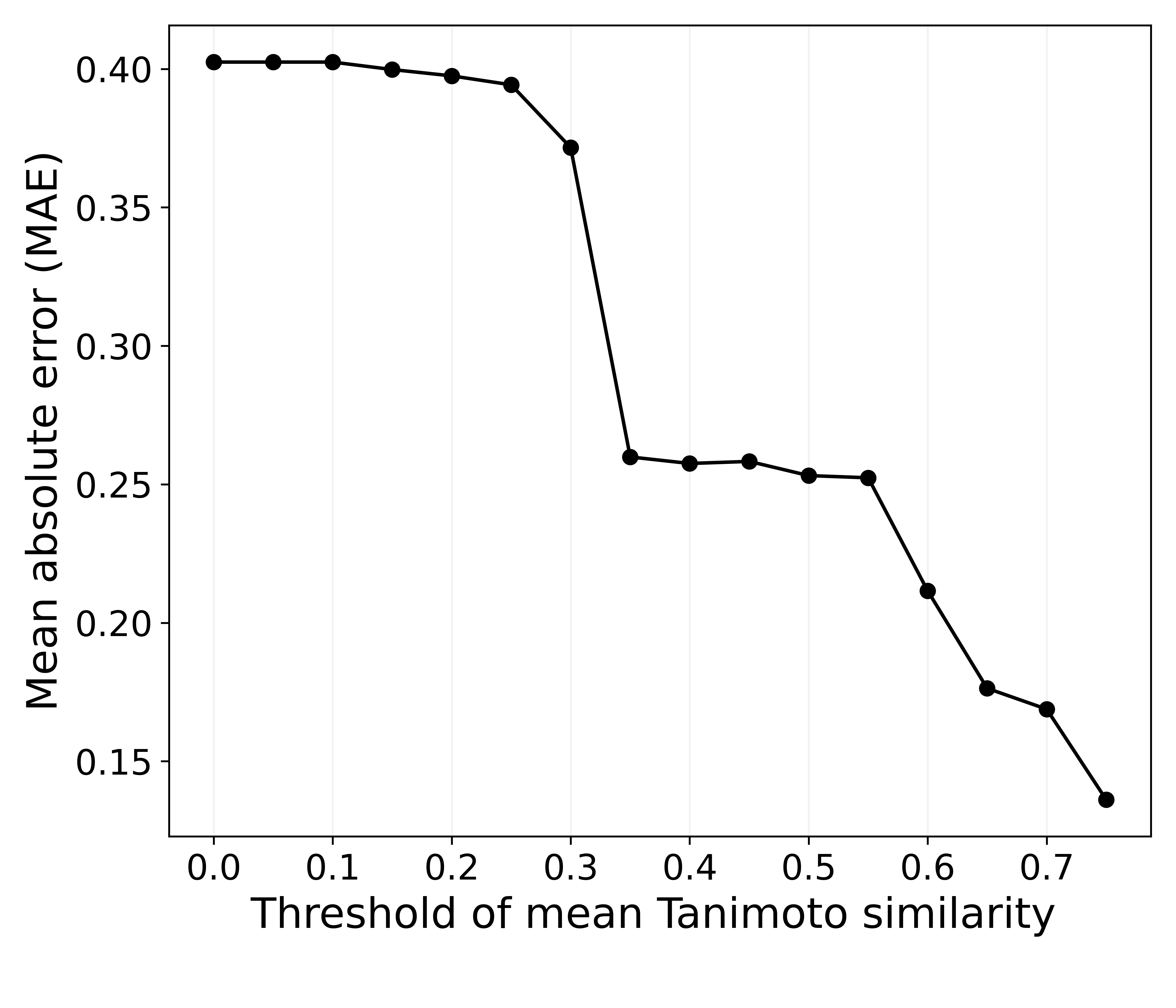}
     \end{subfigure}
    \caption{(Left) Parity plot of the experimental vs. predicted $\ln \gamma_{ij}^\infty$ with the proposed GH-GNN model for (discrete) extrapolation systems with respect to chemical compounds in the external dataset. The gray parity line corresponds to the perfect prediction. The color bar indicates the average Tanimoto similarity between the extrapolated compound and the 10 most similar molecules in the the training set. (Right) Mean absolute error (MAE) achieved by GH-GNN for systems in the external dataset for which the indicated minimum threshold of the Tanimoto similarity indicator is met. This indicator was calculated as the average of the Tanimoto similarities between the extrapolated compound and the 10 most similar molecules in the training set.}
    \label{fgr:parity_discrete_extrapolation_Brouwer_Tanimoto_MAE}
\end{figure}

Figures \ref{fgr:parity_discrete_extrapolation_Brouwer_Nsystems_MAE} (left) and \ref{fgr:parity_discrete_extrapolation_Brouwer_Tanimoto_MAE} (left) show the parity plot of GH-GNN when tested on the external dataset composed of 2233 binary systems. The color bar in Figure  \ref{fgr:parity_discrete_extrapolation_Brouwer_Nsystems_MAE} (left) displays the number of binary-systems in the training set that contain the same combination of solute-solvent chemical classes. It can be observed that well-represented systems in the training set (in terms of chemical classes) tend to be predicted well. This relationship is indeed helpful for investigating the applicability domain of GH-GNN, and a similar analysis can be applied to other predictive models for the same purpose. To illustrate the impact that the number of representative systems in the training set has on the prediction accuracy, Figure \ref{fgr:parity_discrete_extrapolation_Brouwer_Nsystems_MAE} (right) shows the MAE that GH-GNN achieves for systems in the external dataset, for which the specified threshold of number of systems with the same solute-solvent chemical classes are contained in the training set. A MAE of around 0.33 is achieved for systems that have a representation of 25-100 same-class systems in the training set. This level of accuracy is remarkable given the low requirement of having at least 25 same-class systems in the training set. For the external dataset, 1134 systems (corresponding to around 50\% of the total number of systems in the external dataset) have a representation of at least 25 same-class systems. As can be seen from the complete list of binary-classes contained in the training set (available in section S8 of the Supporting Information) used in this work, from the 841 solute-solvent class combinations 105 combinations (indicated with a star in section S8 of the Supporting Information) can be potentially predicted with similar accuracy.  

As presented above, the analysis of chemical classes provides a general and practical way for getting insights into the applicability domain of a $\gamma_{ij}^\infty$ model. However, situations might occur in which a molecular system is well-represented in terms of chemical classes (e.g., by having more than 25 systems with the same solute-solvent chemical classes in the training set), but is still poorly predicted by the model. This is reflected in the fluctuations in performance observed in Figure \ref{fgr:parity_discrete_extrapolation_Brouwer_Nsystems_MAE} (right). For instance, one can have many systems where the solutes are classified as ``saturated hydrocarbons", but all of them are composed by short-chain molecules, and still a system containing a long-chain ``saturated hydrocarbon" could be miss-predicted. The reason is that, even though the corresponding chemical class might be well-represented in terms of the number of systems, the variation of $\ln \gamma_{ij}^\infty$ within the same class might still be high (in the hypothetical example of ``saturated hydrocarbons" likely due to the different entropic contributions to $\gamma_{ij}^\infty$ caused by differences of molecular size).

Hence, in order to provide more reliable indications of the model's applicability domain a measure of the queried system's similarity with the systems in the training set needs to be also provided. For the extrapolation studies with respect to chemical structures presented in this work, the Tanimoto similarity between the extrapolated compound and the chemical species in the training set is used. The color map in Figure \ref{fgr:parity_discrete_extrapolation_Brouwer_Tanimoto_MAE} (left) presents the average Tanimoto similarity between the extrapolated species and the 10 most similar molecules in the training set (referred to as the Tanimoto indicator). A relationship between the Tanimoto indicator and the accuracy of the predictions can be observed, because systems with high Tanimoto indicator values tend to be predicted with higher accuracy. This is confirmed by looking at Figure \ref{fgr:parity_discrete_extrapolation_Brouwer_Tanimoto_MAE} (right) which shows the MAE that GH-GNN achieves for systems in the external dataset for which the indicated threshold of the Tanimoto indicator is met. It can be observed that even when the averaged Tanimoto similarity between the extrapolated compound and the 10 most similar molecules in the training set is 0.35 or higher, the MAE is below 0.3. Having a Tanimoto indicator higher than 0.6 leads to a MAE below 0.2. Similarly to the chemical classes indicator, a relationship between the similarity of the extrapolated species and the quality of the model's prediction is observed. Therefore, it can be stated that GH-GNN achieves a MAE below 0.3 in extrapolation to new chemical species whenever the following two conditions are met:  1) At least 25 systems in the training set have the same solute-solvent chemical classes, and 2) the extrapolated molecule has a Tanimoto indicator higher than 0.35. 

\section{Conclusions}
In order to develop more sustainable chemical processes in general, and separation processes in particular, the accurate prediction of physicochemical properties of mixtures, such as $\gamma_{ij}^\infty$, is of paramount importance. In this work, we have first studied the performance of previously proposed GNN-based models \citep{D1DD00037C, qin2022capturing} by comparing them to the popular UNIFAC-Dortmund, MOSCED \citep{lazzaroni2005revision} and COSMO-RS models for predicting $\ln \gamma_{ij}^\infty$. In general, GNN-based models outperform the studied mechanistic models in terms of the MAE. Then, we have developed a graph neural network framework that utilizes a simple Gibbs-Helmholtz-derived expression for capturing the temperature dependency of $\ln \gamma_{ij}^\infty$. We call this model Gibbs-Helmholtz Graph Neural Network (GH-GNN) in order to highlight the hybrid arrangement of this model by using a data-driven approach together with a simple, but solid thermodynamics-based relationship for capturing the temperature dependency of $\ln \gamma_{ij}^\infty$. Moreover, GH-GNN makes use of global-level molecular descriptors that capture the polarizability and polarity of the solutes and solvents involved which are related to ``dipole - dipole", ``dipole - induced dipole" and ``induced dipole - induced dipole" interactions that affect the Gibbsian excess enthalpy and thus the value of $\ln \gamma_{ij}^\infty$. These descriptors were inspired by the advantages of including explicit information about intermolecular interactions into the modeling framework similar to the MOSCED and SolvGNN \citep{qin2022capturing} models. GH-GNN achieves better results than the UNIFAC-Dortmund method overall. Moreover, GH-GNN is able to predict systems that UNIFAC-Dortmund is simply not able to predict due to the lack of the required binary-interaction parameters. This results in a much more flexible and accurate framework that can be exploited for calculating $\ln \gamma_{ij}^\infty$ values for systems of practical relevance ranging from solvent screening to environmental studies.

In addition, a series of inter-/extrapolation studies regarding temperature (continous) and chemical compounds (discrete) are presented to analyze the performance and applicability domain of GH-GNN. Overall, the proposed GH-GNN model is able to inter-/extrapolate to different system temperatures with great accuracy given that the Gibbs-Helmholtz-derived equation captures most of this dependency. By contrast, the accuracy of the GH-GNN's predictions is mostly related to the type of chemical compounds involved in the liquid mixture of interest. We studied the inter-/extrapolation capabilities of GH-GNN to different solute-solvent combinations and gave indications about the applicability domain and expected accuracy of this model. GH-GNN can predict $\ln \gamma_{ij}^\infty$ with high accuracy when interpolating in the training solute-solvent matrix and when extrapolating to solvent or solute species which are well-represented in the training set in terms of chemical classes (at least 25 solute-solvent systems with the same chemical class should be present in the training set) and Tanimoto indicator (higher than 0.35). With this we also hope to highlight the importance of a careful data splitting and the relevance of providing practical indications on the applicability domain of data-driven models.  The analysis regarding the applicability domain of multi-component systems should be extended and further investigated in directions such as the definition of multi-component similarity metrics and the reliability of distance metrics in the GNN-embedding space. The GH-GNN framework can also be extended to capture the composition dependency of $\ln \gamma_{ij}$ and can be potentially transferred to the prediction of fugacity coefficients.

\bibliographystyle{unsrtnat}
\bibliography{references}  

\end{document}